%% file: main.tex
\documentclass[sigconf]{acmart}
\setcopyright{none}
\renewcommand\footnotetextcopyrightpermission[1]{}

\AtBeginDocument{%
  }

\setcopyright{acmlicensed}
\copyrightyear{2018}
\acmYear{2018}
\acmDOI{XXXXXXX.XXXXXXX}
\acmConference{ACM MobiHoc}{November 23-26, 2026}{Tokyo Japan}
\acmISBN{978-1-4503-XXXX-X/2018/06}

\usepackage{xcolor}
\usepackage{amsfonts}
\usepackage{balance}
\PassOptionsToPackage{hyphens}{url}
\usepackage{color}
\usepackage{graphics}
\usepackage{listings}
\usepackage{multicol}
\usepackage{multirow}
\usepackage[scaled]{helvet}
\usepackage{rotating}
\usepackage{xspace}
\usepackage{algorithm}
\usepackage{comment}
\usepackage{enumitem}
\usepackage{amsmath}
\usepackage{mathrsfs}
\usepackage{cancel}
\usepackage{subcaption}
\usepackage{commath}
\usepackage{hyperref}
\usepackage{cleveref}
\usepackage{url}
\usepackage{xurl} 
\usepackage{soul}
\usepackage{booktabs}   

\usepackage{amssymb}    
\usepackage{graphicx}   
\usepackage{caption}    
\usepackage{pifont}     
\usepackage[utf8]{inputenc}
\usepackage[normalem]{ulem}
\usepackage[font=small]{caption}
\usepackage{refcount}

\newcommand{\cmark}{\ding{51}}  
\newcommand{\xmark}{\ding{55}}  
\newcommand{\newtodo}[1]{\ClassWarning{NOT READY TO SUBMIT}{There is something left todo}\textcolor{blue}{}}

\newcommand{\review}[1]{\ClassWarning{NOT READY TO SUBMIT}{There is something left todo} \textcolor{orange}{}}

\newcommand{\shortname}{\textit{FreeChirp}\xspace}

\newcommand{\algoname}{{FSMA}\xspace}

\newcommand{\squishlist}
{
    \begin{list}{$\bullet$}
    {
        \setlength{\itemsep}{0pt}      \setlength{\parsep}{2pt}
        \setlength{\topsep}{2pt}       \setlength{\partopsep}{0pt}
        \setlength{\leftmargin}{1em} \setlength{\labelwidth}{0.5em}
        \setlength{\labelsep}{0.5em}
    }
}
\newcommand{\squishend}
{
    \end{list}
}

\begin{document}



\title{\textit{FSMA}: Gateway-Controlled Access for Scalable LoRa Non-Terrestrial Networks}

\author{Rohith Reddy Vennam$^{\dagger}$, Maiyun Zhang$^{\dagger}$,
Raghav Subbaraman$^{\dagger}$, Deepak Vasisht$^{\S}$, Dinesh Bharadia$^{\dagger}$}
\affiliation{%
  $^{\dagger}$University of California San Diego, La Jolla, CA, USA \\
  $^{\S}$University of Illinois Urbana-Champaign, Urbana, IL, USA \\
  $^{\dagger}$ {\{rvennam, maz005, rsubbara, dineshb\}@ucsd.edu}, $^{\S}$ \{deepakv\}@illinois.edu \\
  \vspace{0.02\textwidth}
  \country{}
}
\renewcommand{\shortauthors}{Vennam et al.}

\input{0_abstract}
\maketitle

\input{1_introduction}

\input{2_backgroud_and_related}

\input{3_design}
\input{4_implementation}

\input{5_evaluation}

\input{6_conclusion}

\bibliographystyle{unsrt}
\bibliography{custom_references}
\label{lastpage}

\end{document}

%% file: 0_abstract.tex

\begin{abstract}

LoRa-based non-terrestrial networks are rapidly becoming infrastructure for global IoT, with LEO satellites delivering direct-to-device connectivity and drone-mounted gateways serving precision agriculture, emergency response, and environmental monitoring. Both settings share a fundamental MAC-layer bottleneck: large coverage footprints cause thousands of devices to collide on the same channel, while gateway mobility induces rapid link variations, making link-unaware transmissions a persistent source of wasted airtime. These challenges demand random access solutions, yet ALOHA provides no collision coordination, CSMA sensing fails beyond ~$15~km$, and BSMA's continuous busy-tone consumes 80--90\% of channel capacity. We propose Free Signal Multiple Access (\textit{FSMA}), a synchronization-free, gateway-controlled MAC protocol. Like a traffic signal, \textit{FSMA} uses a single LoRa upchirp (\textit{FreeChirp}) to indicate channel availability. By transmitting it at a lower spreading factor than the node uplink, \textit{FSMA} implicitly filters weak-link nodes through channel reciprocity, reducing the collision window by 6--200$\times$ without synchronization, handshakes, or hardware modifications. Hardware experiments with 25 COTS LoRa nodes and a drone-mounted gateway demonstrate 2$\times$ higher throughput, 2.5--5$\times$ better packet reception ratio, and up to 5$\times$ improved energy efficiency. Large-scale simulations with real TLE-based satellite trajectories confirm scalability to 5000+ devices per satellite pass.

\end{abstract}

%% file: 1_introduction.tex
 \vspace{-0.02\textwidth}
\section{Introduction}\label{sec:intro}

\begin{figure}
     \centering
     \begin{subfigure}[b]{0.20\textwidth}
         \centering
         \includegraphics[width=\textwidth]{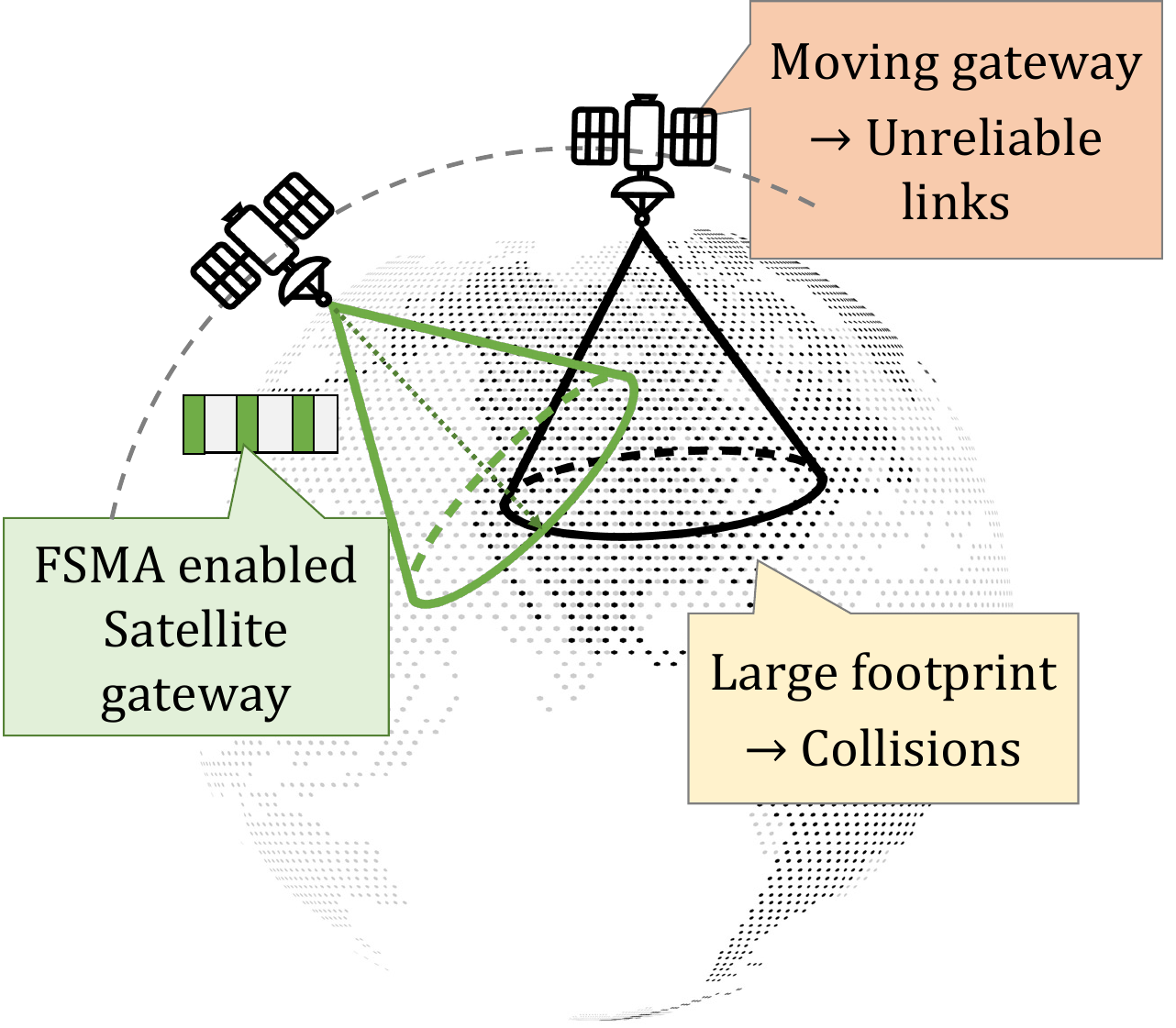}
         \caption{\centering Direct-to-Satellite IoT \\(Ubiquitous coverage).}
         \label{fig:satellite_motivation}
     \end{subfigure}
     \hspace{0.01\textwidth}
     \begin{subfigure}[b]{0.23\textwidth}
         \centering
         \includegraphics[width=\textwidth]{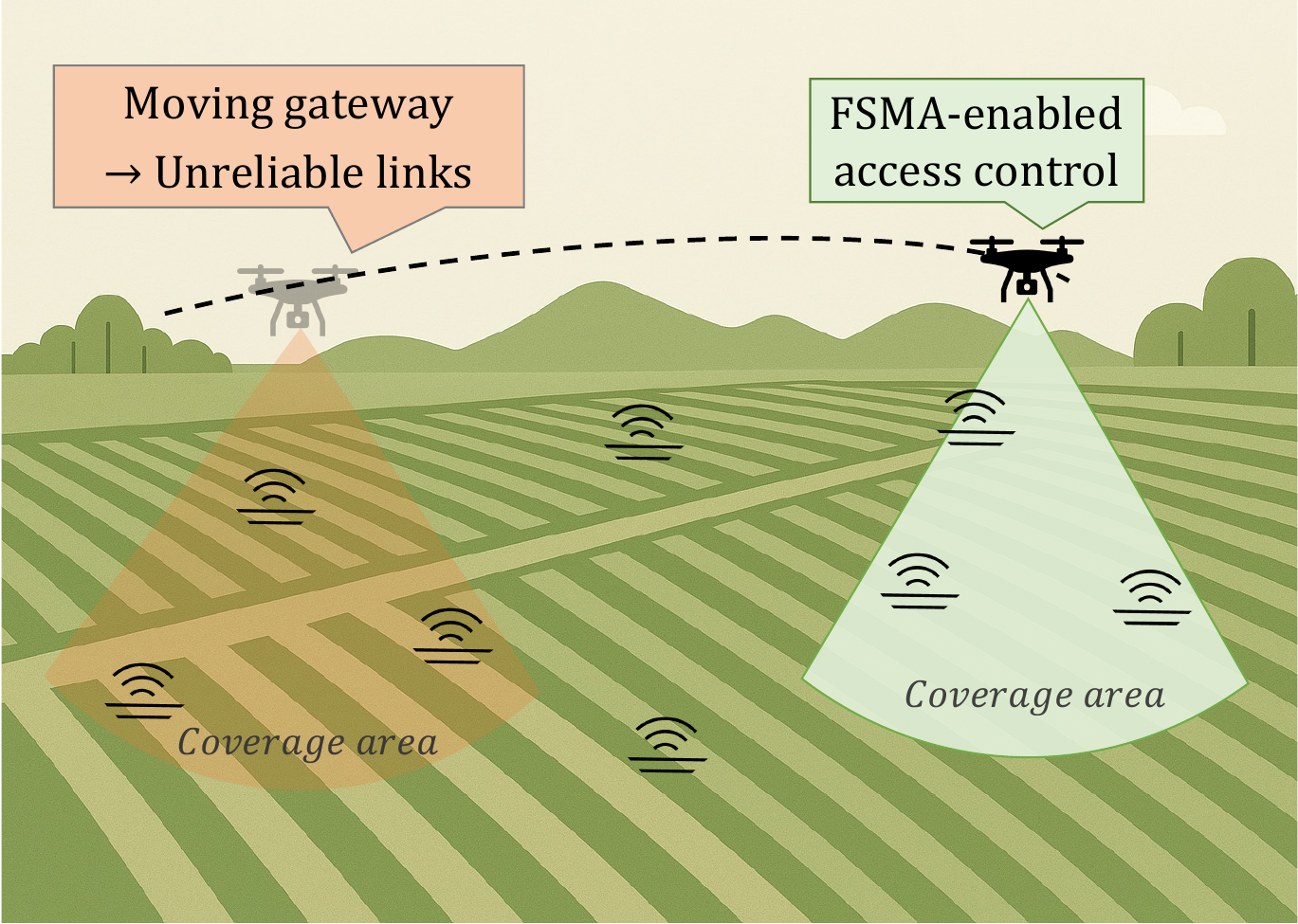}
         \caption{\centering Drone-based IoT scenario \\(Precision farming).}
         \label{fig:drone_motivation}
     \end{subfigure}
     \setlength{\belowcaptionskip}{-4pt}
     \vspace{0.01\textwidth}
     \caption{\textit{Non-terrestrial LoRa networks experience packet
     loss from both collisions and link failures.} In~(a), a satellite's
     vast footprint causes numerous devices to contend for the same
     channel, leading to collisions. In both~(a) and~(b), gateway
     mobility creates dynamic links that alternate between high SNR,
     low SNR, and outage phases. FSMA addresses both challenges through
     gateway-controlled, link-aware channel access.}
     \label{fig:motivation}
     \vspace{-0.01\textwidth}
\end{figure}

LoRa-based non-terrestrial networks (NTNs) are an active and growing
infrastructure for global IoT connectivity. Satellite constellations
including FOSSA, Swarm, Echostar, and
Tianqi~\cite{fossa,semtech_swarm_lora,tinyGS}
already deliver direct-to-satellite LoRa links, while drone-mounted
gateways serve precision agriculture, emergency response, and
environmental monitoring~\cite{dji_agri_drone}.
Despite this momentum, both settings share an unresolved bottleneck at
the MAC layer: \textbf{packet loss from collisions and unreliable links}.

\begin{figure*}[t!]
     \centering
     \begin{subfigure}[t]{0.24\textwidth}
         \centering
         \includegraphics[width=\textwidth]{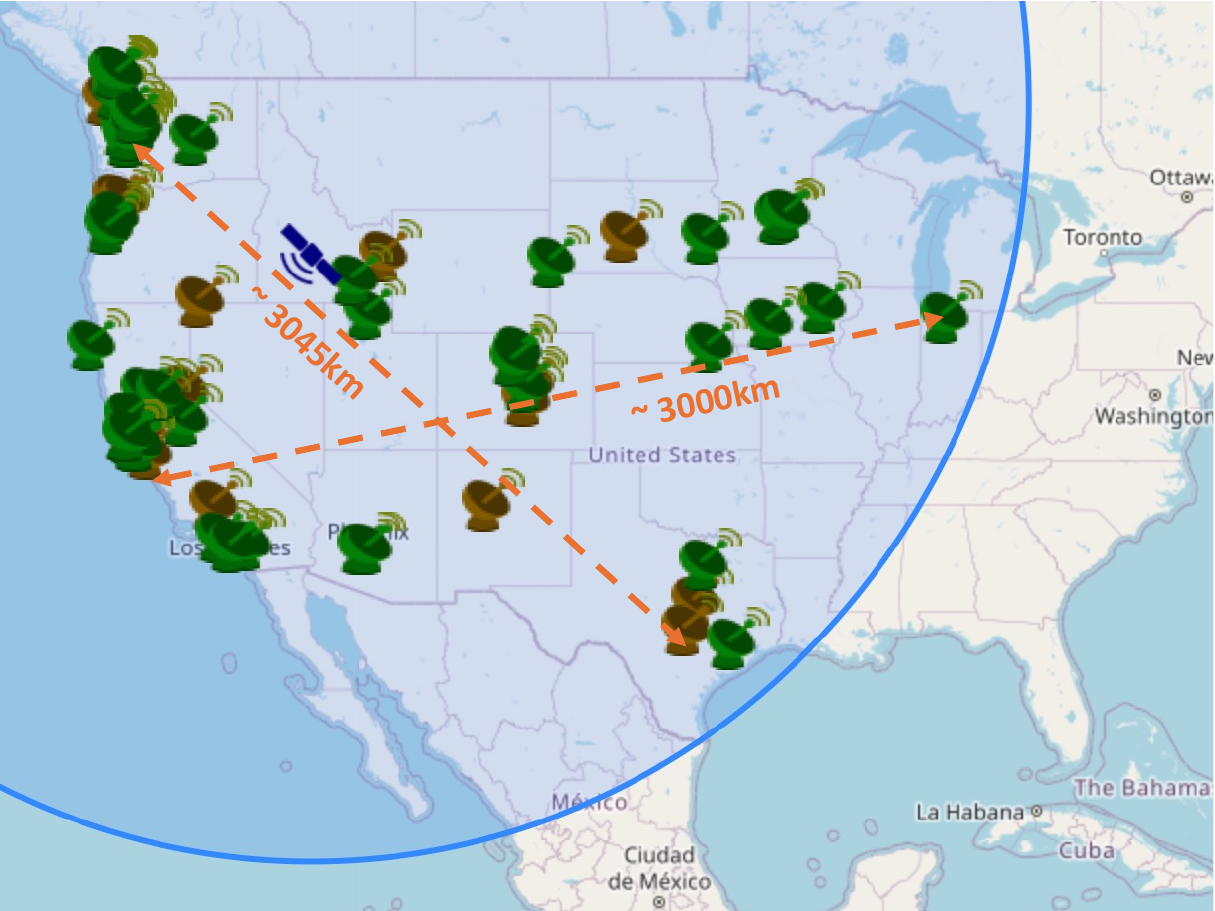}
         \caption{Ground stations receiving the same Norby-2 packet
         ($>$3000~km coverage).}
         \label{fig:intro_norby2_coverage}
     \end{subfigure}
     \hfill
     \begin{subfigure}[t]{0.24\textwidth}
         \centering
         \includegraphics[width=\textwidth]{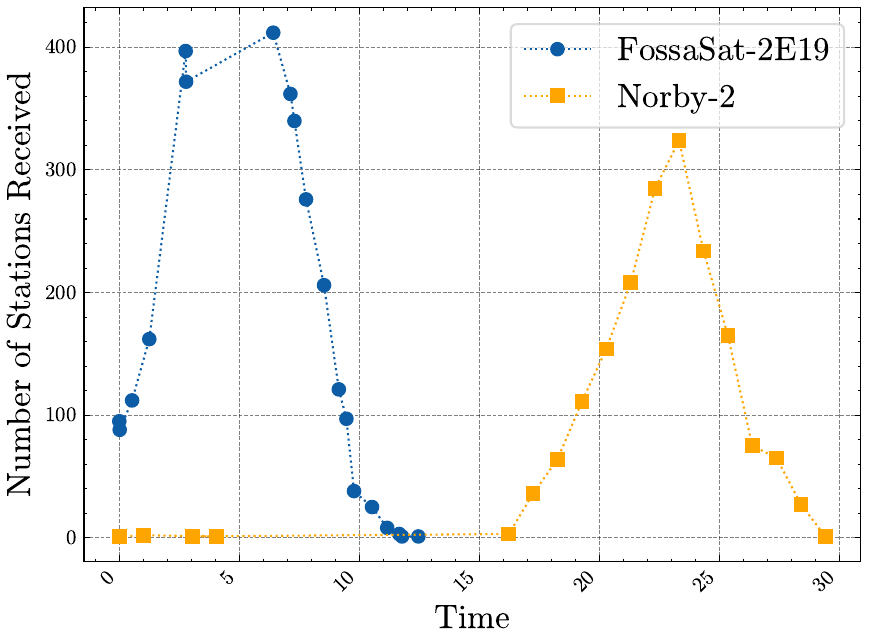}
         \caption{Active ground stations per packet: Norby and Fossa
         satellites over 30~minutes.}
         \label{fig:intro_num_ground_stations_received}
     \end{subfigure}
     \hfill
     \begin{subfigure}[t]{0.24\textwidth}
         \centering
         \includegraphics[width=\textwidth]{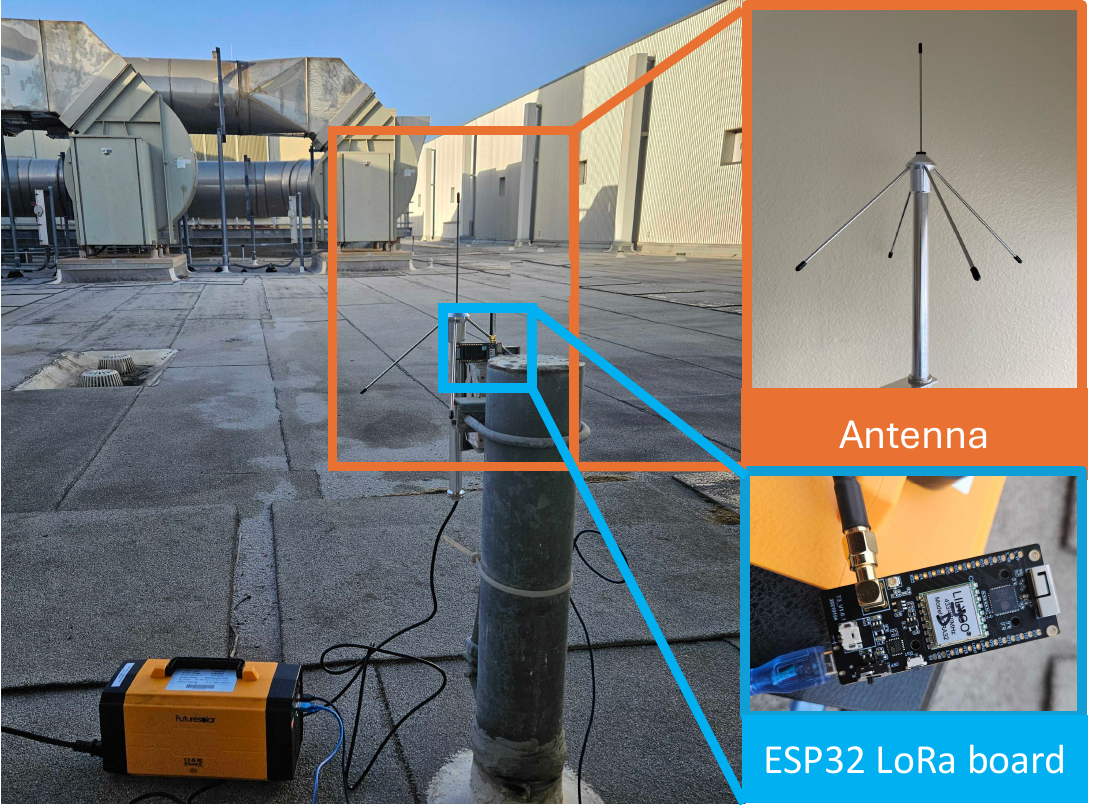}
         \caption{TinyGS ground station built with an ESP32 board and
         433~MHz antenna.}
         \label{fig:intro_tinygs_groundstation_setup}
     \end{subfigure}
     \hfill
     \begin{subfigure}[t]{0.24\textwidth}
         \centering
         \includegraphics[width=\textwidth]{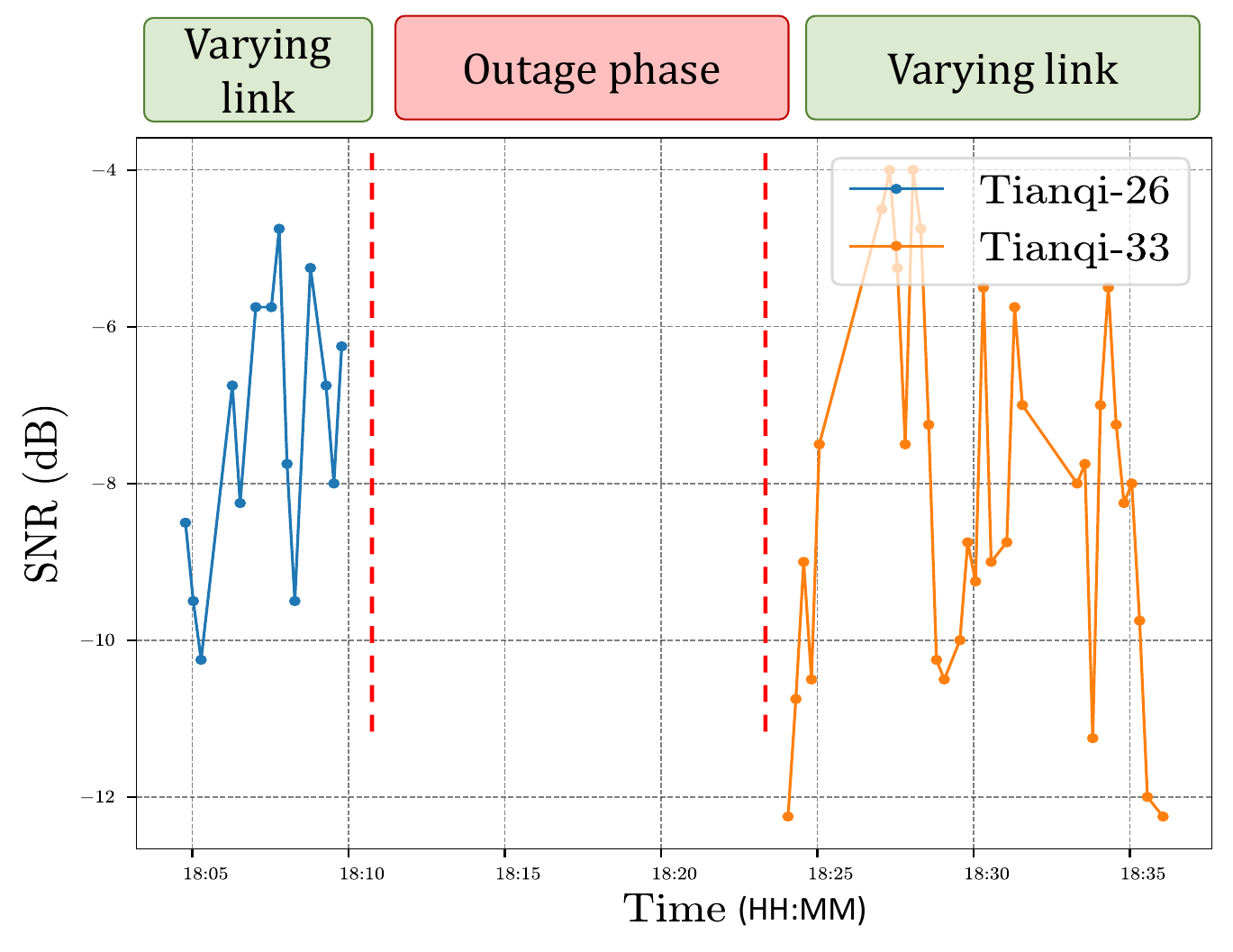}
         \caption{SNR from Tianqi-26 and Tianqi-33 over a 30-minute
         window: up to 10~dB variation with complete outage phases.}
         \label{fig:intro_tinygs_link_snr_analysis}
     \end{subfigure}
    \caption{Quantifying the large-footprint and dynamic-link challenges
    using real satellite packet traces from TinyGS ground
    stations~\cite{tinyGS, tinyGSNorby2Packet2025April22}.
    (a)~A single packet covers over 3000~km; (b)~active stations drop
    by over 100$\times$ within five minutes; (d)~SNR varies by up to
    10~dB with complete outage phases, even within a continuous
    visibility window.}
    \label{fig:tinygs_trace_analysis}
\end{figure*}

Real-world measurements quantify the severity of both problems
(Figure~\ref{fig:tinygs_trace_analysis}). A single LoRa packet from
Norby-2 was simultaneously received by ground stations spanning over
3000~km across North America---from Seattle to Houston and San
Francisco to Chicago---indicating that thousands of devices contend on
the same channel concurrently
(Figure~\ref{fig:intro_norby2_coverage}). Active stations receiving
Fossa packets dropped from 412 to fewer than 3 within five minutes as
the satellite moved, reflecting how rapidly the contending population
changes (Figure~\ref{fig:intro_num_ground_stations_received}). Even
within a continuous satellite visibility window, measured SNR
fluctuates by up to 10~dB and passes through complete outage phases
(Figure~\ref{fig:intro_tinygs_link_snr_analysis}). These observations
establish two root causes of packet loss in NTN LoRa:

\begin{itemize}

\item \textbf{\textit{Large footprint causing collisions:}}
A single satellite covers over 3000~km in diameter, allowing channel access to thousands
of devices simultaneously. Terrestrial LoRa
gateways cover a few kilometers; NTN gateways cover orders of magnitude
more, making uncoordinated channel access fundamentally untenable at
this scale.

\item \textbf{\textit{Mobile gateway causing unreliable links:}}
Satellites and drones are constantly in motion, producing short,
shifting visibility windows and rapidly varying link quality. A
transmission viable at one moment fails minutes later as the gateway
geometry changes, making link-unaware access a persistent source of
wasted airtime and packet loss.

\end{itemize}

Addressing both challenges simultaneously is fundamentally hard.
Node-side channel sensing covers only 5--15~km~\cite{subbaraman2022bsma},
less than 0.05\% of a typical satellite footprint, so any useful
coordination must originate from the gateway. Yet the gateway must
coordinate access and provide link quality feedback without consuming
the channel it is trying to free, without requiring time
synchronization across a continuously changing node population, and
without hardware changes to existing LoRa devices. No existing protocol
satisfies all of these constraints together.

Existing MAC protocols each address only a subset of these requirements
(Table~\ref{tab:mac_comparison}). The large coverage footprint, rapidly
changing active nodes and gateway mobility together rule out an
entire class of solutions: synchronization-based protocols require
nodes to coordinate timing with a moving gateway whose active coverage
changes continuously, while RTS/CTS-style schemes incur handshake
overhead that exceeds 200\% of the payload size and collapses under
simultaneous RTS collisions at high node densities~\cite{ortigueira2021ress,
o2020practical}. These constraints mandate random access solutions that
require no back-and-forth exchanges between nodes and the gateway.
Within the random access class, however, every existing approach fails
in a different way.
\textit{ALOHA}~\cite{abramson1970aloha, vogelgesang2021uplink,
shenoy2024cosmac} requires no synchronization and runs on COTS hardware,
but provides no collision coordination: collision rates grow with node
density, and the protocol has no mechanism to suppress transmissions
over weak or unavailable links. \textit{CSMA}~\cite{gamage2023lmac,
pham2021dense} reduces collisions through channel sensing, but its CAD
range is limited to 3--15~km~\cite{subbaraman2022bsma}. At satellite
scale, over 99.9\% of concurrently active nodes lie outside any single
node's sensing range, rendering CSMA equivalent to ALOHA over
NTN footprints. \textit{Slotted ALOHA and its
variants}~\cite{polonelli2018slotted, trub2018increasing,
piyare2018demand} reduce collisions through fixed time slots but
require strict synchronization, which is impractical when the gateway
is moving and the active node population changes continuously.
\textit{Busy-tone multiple access
(BSMA)}~\cite{subbaraman2022bsma, tobagi1975packet, wu1987receiver,
haas2002dual} overcomes the sensing range limitation by having the
gateway broadcast a busy tone, but this continuous transmission occupies
the channel for 80--90\% of the time under high load and requires either
a dedicated frequency channel or a full-duplex gateway, neither of which
is practical in energy-constrained NTN deployments. Critically, none
of these protocols support link-aware transmissions: all assume that a
collision-free packet will be reliably received, an assumption that
fails under gateway mobility.

\begin{figure}[!t]
    \begin{minipage}[t]{0.49\textwidth}
    \centering
     \begin{subfigure}[t]{0.48\textwidth}
         \centering
         \includegraphics[height=0.165\textheight]{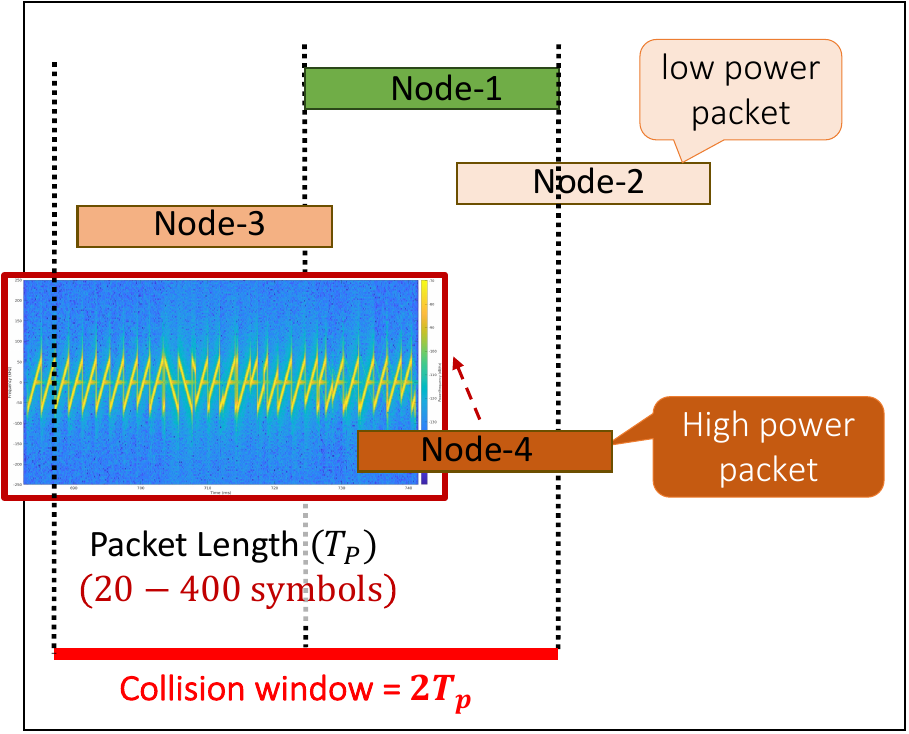}
         \caption{\centering Baselines (ALOHA, CSMA)}
         \label{fig:baseline_collision_space}
     \end{subfigure}
     \hfill
     \begin{subfigure}[t]{0.48\textwidth}
         \centering
         \includegraphics[height=0.165\textheight]{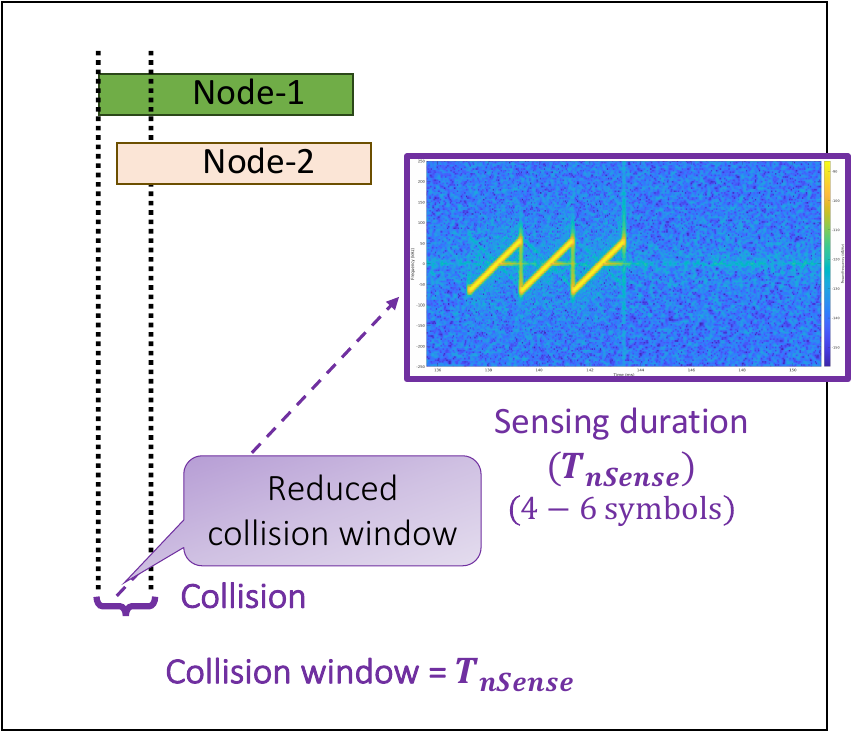}
         \caption{\centering \algoname}
         \label{fig:fsma_collision_space}
     \end{subfigure}
     \caption{Collision window comparison. In baseline LoRa (ALOHA,
     CSMA), collisions span the 2 $\times$ packet duration (20--400~symbols).
     \algoname confines contention to the sensing window
     ($<$6~symbols), reducing the collision window by
     6--200$\times$ depending on packet length.}
     \label{fig:collisions}
     \end{minipage}
     \vspace{-0.02\textwidth}
\end{figure}

\textbf{\algoname}: We introduce Free Signal Multiple Access (\algoname), a novel synchronization-free, gateway-controlled MAC protocol for LoRa networks that significantly reduces collisions and enables link-aware transmissions. 
FSMA acts like a traffic light: it allows nodes to transmit when the channel is free and forces them to back off when the channel is busy. It uses a single LoRa upchirp (/) as a free signal (\shortname) to coordinate channel access. When the channel is idle, the gateway transmits a \shortname and waits for a short interval. Nodes with data wake up and listen for this signal; upon detecting a \shortname, a node transmits its packet, otherwise it backs off and retries after a random delay. Once the gateway detects a transmission, it stops sending {\shortname}s until the channel becomes free again, effectively coordinating access without synchronization or control packet exchange. It further reduces the collision window and addresses the dynamic link behavior as follows: 

\begin{figure}[t]
     \begin{minipage}[t]{0.48\textwidth}
         \centering
         \includegraphics[width=0.7\textwidth]{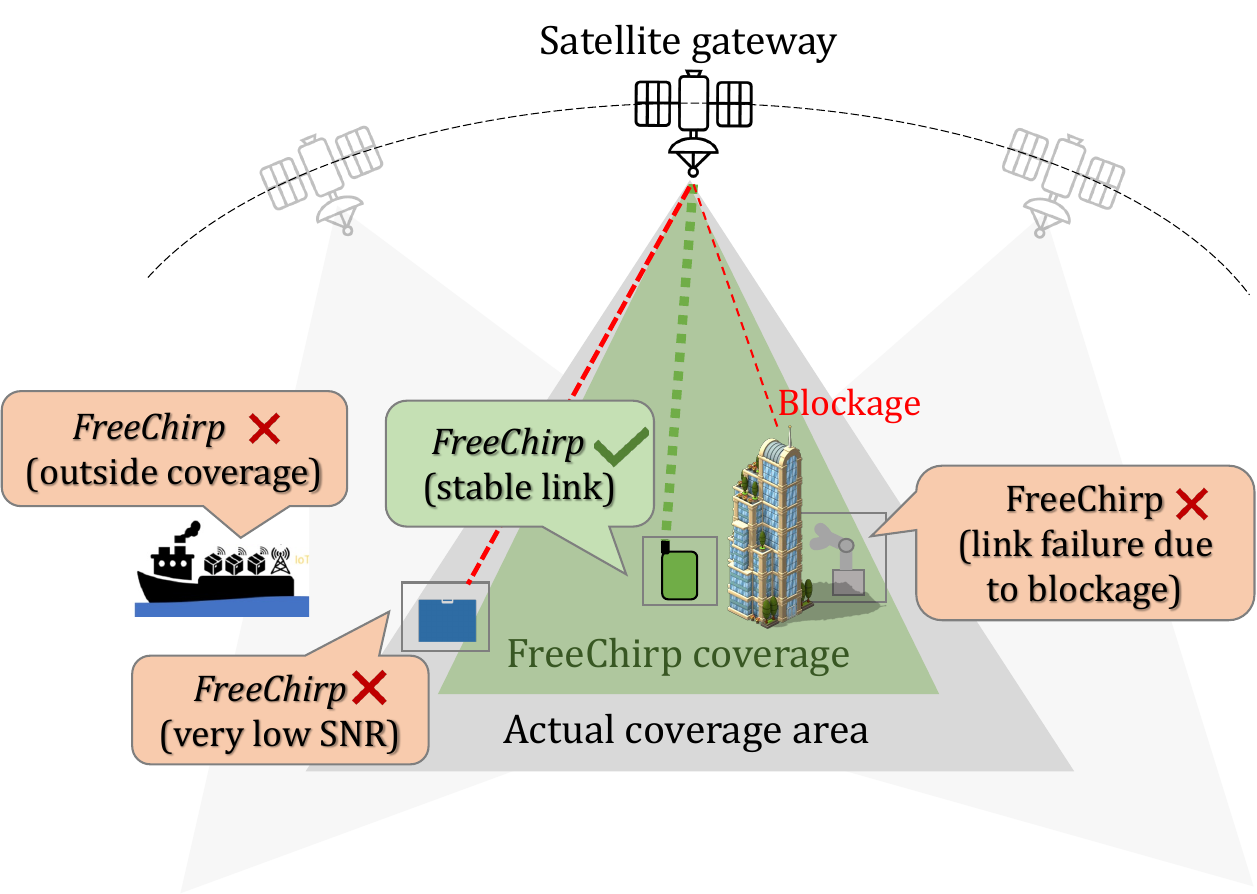}
          \vspace{-0.04\textwidth}
        \caption{
        \shortname coverage and link-aware filtering in
        \algoname. A lower-SF \shortname implicitly filters weak-link
        nodes through channel reciprocity; only nodes with strong uplinks
        detect and respond. Gateway mobility continuously shifts coverage,
        providing long-term fairness without explicit scheduling.
                }
        \label{fig:fsma_beacons}
     \end{minipage}
    \vspace{-0.01\textwidth}
\end{figure}

\textbf{\textit{1) Reducing the collision window:}} We define the collision 
window as the interval during which simultaneous transmissions from multiple 
nodes occur, leading to packet loss at the gateway. In conventional LoRa 
networks (Figure~\ref{fig:baseline_collision_space}), any transmission 
overlapping an ongoing reception causes a collision: a node that begins 
transmitting up to $T_P$ before the current packet creates a tail-overlap 
collision, and a node that begins up to $T_P$ after creates a head-overlap 
collision. Both directions contribute, making the collision window equal to 
twice the packet duration: $T_{cw}^{\text{ALOHA}} = 2T_P$, spanning 40.5 
to 808.5 symbols for payloads of 0 to 192 
bytes~\cite{wiki_swarm,lora_calculator}. FSMA ensures that a new 
\shortname is only sent after confirming no transmission was triggered 
by the previous signal, thereby restricting collisions to nodes that detect 
the same \shortname, reducing the collision window to the node sensing 
period (Figure~\ref{fig:fsma_collision_space}), typically fewer than 6 node 
symbols. A natural follow-up question is: what happens if many nodes detect 
the same \shortname and attempt to transmit simultaneously? If all 
nodes continuously sense for \shortname, simultaneous detections still 
cause collisions. To address this, \algoname introduces a random exponential 
backoff mechanism that spreads transmission attempts across time, reducing 
contention within the sensing window.

\textbf{\textit{2) Enabling link-aware access:}}
In terrestrial deployments, a collision-free transmission is decoded 
reliably: gateways are static and links are stable. In NTN scenarios,
this assumption breaks. Link quality fluctuates by more than 10~dB
within a single satellite pass and drops to a complete outage between
passes (Figure~\ref{fig:intro_tinygs_link_snr_analysis}), making
link-unaware transmissions a persistent source of wasted airtime even
when collisions are avoided.
FSMA addresses this through channel reciprocity. The gateway transmits
\shortname at a lower SF than the node uplink: a node that detects the
downlink chirp confirms both channel availability and adequate link
quality for its higher-SF uplink. Nodes with weak links, blockage, or
outside instantaneous coverage receive no \shortname and apply backoff,
avoiding futile transmissions. Transmitting at a lower SF also reduces
the \shortname's ground coverage, further limiting the number of
competing nodes.
This raises a fairness concern: nodes with consistently weak links risk
being permanently excluded. However, unlike static terrestrial gateways,
a moving satellite or drone continuously shifts its coverage footprint.
A node excluded at one gateway position gains access as the geometry
changes, providing inherent long-term fairness across all devices
without any need for explicit fairness mechanism.

\textbf{\textit{Evaluation Overview:}}
FSMA requires no custom hardware. At nodes, existing LoRa deployments
require only a firmware update to enable \shortname detection. At the
gateway, FSMA is realized by pairing two off-the-shelf LoRa
modules, one for packet reception and the other for \shortname
transmission, without any hardware modifications. We prototype FSMA on
COTS SX1272/76-based nodes (Semtech mBed and Adafruit Feather) and
evaluate it in static indoor and dynamic outdoor settings.
To replicate high collision rates and dynamic link conditions, we
deploy a drone-mounted gateway with 25 nodes distributed across a
campus-scale area. FSMA achieves up to 2$\times$ higher throughput,
2.5--5$\times$ improvement in packet reception ratio, and up to
5$\times$ lower energy per successful packet compared to baselines,
including sensing overhead. For large-scale validation, we develop
\textit{NTNLoRa}, a custom Python-based PHY-MAC simulator with real TLE-based
satellite trajectories and end-to-end system modeling. Using a
Starlink satellite pass with nodes distributed across western North
America, \textit{NTNLoRa} demonstrates that FSMA sustains these gains while
scaling to 5000+ devices within a 10-minute satellite visibility
window at a 0.1\% duty cycle.

\noindent
\textbf{Contributions:} We summarize our contributions as follows
\vspace{-0.005\textheight}
\begin{itemize}
    \item \textbf{MAC Protocol Design:} We introduce \algoname, a 
    gateway-controlled MAC protocol in which a single LoRa upchirp 
    (\shortname `/') coordinates channel access across a 
    large footprint, without synchronization, handshakes, or a 
    dedicated control channel, reducing the collision window by 
    6$\times$--200$\times$ over ALOHA and CSMA with negligible gateway 
    overhead.
    
    \item \textbf{Link-Aware Access via Channel Reciprocity:} We
    demonstrate that transmitting FreeChirp at a lower SF than the node
    uplink silently filters nodes with weak or unreliable links through
    channel reciprocity, enabling link-aware access without any additional 
    signaling, a feature that no existing LoRa MAC protocol provides.
    
    \item \textbf{Implementation and Validation:} FSMA deploys on
    existing LoRa hardware via a firmware-only node update and two
    paired COTS gateway modules. Hardware experiments show 2$\times$
    throughput, 2.5--5$\times$ packet reception ratio, and up to 5$\times$
    energy efficiency gains. \textit{NTNLoRa} simulations confirm scalability to
    5000+ devices per satellite pass.
    

    \item \textbf{Open-Source Release:} We make the \textit{NTNLoRa} simulator and complete \algoname implementation publicly available at \href{https://github.com/ucsdwcsng/fsma-lora}{\underline{github.com/ucsdwcsng/fsma-lora}}, enabling reproducibility and further research on NTN LoRa MAC protocols.
\end{itemize}

%% file: 2_backgroud_and_related.tex
\vspace{-0.01\textheight}
\section{Background and Related Work} \label{sec:background}

\subsection{Case Study with Real-World Satellite Packets}
\label{sec:casestudy}

To quantify the two challenges introduced in Section~\ref{sec:intro},
we analyze packet traces from operational satellites received by TinyGS
ground stations~\cite{tinyGS} and custom-built receivers
(Figure~\ref{fig:intro_tinygs_groundstation_setup}), characterizing
(i)~spatial coverage extent, (ii)~temporal variation in the active
node population, and (iii)~link SNR fluctuation over a satellite pass.

\begin{itemize}
\vspace{-0.005\textheight}

\item \textbf{Large coverage footprint:}
Figure~\ref{fig:intro_norby2_coverage} shows a single LoRa packet from
Norby-2 received simultaneously by ground stations spanning over
3000~km, from Seattle to Houston and San Francisco to Chicago. With
thousands of IoT devices already deployed in such footprints and
deployments continuing to grow, efficient channel coordination at
continental scale is not optional.

\item \textbf{Dynamic network load:}
Figure~\ref{fig:intro_num_ground_stations_received} shows active TinyGS
stations receiving the same Norby and Fossa packets within a 30-minute
window. Stations dropped from 412 to fewer than 3 within five minutes, 
$>$100$\times$ reduction, as the satellite traversed its orbit. This
rapid change in the contending population rules out synchronization and
handshake-style approaches, which cannot track a node set that changes
on a per-minute timescale.

\item \textbf{Rapid link variation:}
Figure~\ref{fig:intro_tinygs_link_snr_analysis} shows SNR at a fixed
ground receiver from Tianqi-26 and Tianqi-33 over 30~minutes. Link
quality varied by up to 10~dB within a single visibility window and
dropped to complete outage phases between passes, confirming that
link-unaware transmissions waste airtime even when collisions are
absent.

\end{itemize}
\vspace{-0.01\textwidth}

These measurements jointly establish that an effective NTN LoRa MAC
protocol must coordinate access at a large footprint, adapt to rapidly
varying link conditions, and operate without synchronization or
pre-computed transmission schedules.

\begin{table*}[ht]
    \centering
    \small
    \setlength{\tabcolsep}{5pt}
    \caption{Comparison of MAC protocols across key requirements for
    mobile-gateway NTN IoT deployments. Link-aware access is the only
    requirement no prior protocol satisfies; FSMA is the only protocol
    satisfying all requirements simultaneously.}
    \begin{tabular}{l|ccccccc}
        \toprule
        \textbf{Requirement}
            & \textbf{ALOHA}
            & \textbf{CSMA}
            & \textbf{BSMA}
            & \textbf{Slotted}
            & \textbf{RTS/CTS}
            & \textbf{CosMac}~\cite{shenoy2024cosmac}/
            & \textcolor{blue}{\textbf{FSMA}} \\
        \textbf{}
            & \cite{abramson1970aloha,vogelgesang2021uplink}
            & \cite{gamage2023lmac,pham2021dense}
            & \cite{subbaraman2022bsma}
            & \textbf{ALOHA}~\cite{polonelli2018slotted}
            & \cite{ortigueira2021ress, o2020practical}
            & \textbf{SaterIoT}~\cite{ren2024sateriot}
            & \textbf{(This Work)} \\
        \midrule
        Collision Mitigation
            & \xmark & \xmark & \cmark & \cmark & \cmark & \cmark
            & \textbf{\cmark} \\
        Link-Aware Access
            & \xmark & \xmark & \xmark & \xmark & \xmark & \xmark
            & \textbf{\cmark} \\
        Synchronization-Free
            & \cmark & \cmark & \cmark & \xmark & \cmark & \xmark
            & \textbf{\cmark} \\
        NTN Footprint Support
            & \xmark & \xmark & \cmark & \xmark & \xmark & \cmark
            & \textbf{\cmark} \\
        Node Energy Efficiency
            & \xmark & \xmark & \cmark & \xmark & \xmark & \xmark
            & \textbf{\cmark} \\
        Gateway Energy Efficiency
            & \cmark & \cmark & \xmark & \cmark & \cmark & \cmark
            & \textbf{\cmark} \\
        COTS Hardware Compatible
            & \cmark & \cmark & \xmark & \cmark & \cmark & \xmark
            & \textbf{\cmark} \\
        \bottomrule
    \end{tabular}
    \label{tab:mac_comparison}
\end{table*}

\subsection{MAC Protocols for LoRa NTN}
\label{sec:mac_background}
 
We survey prior MAC protocols in two tiers: direct baselines evaluated
in Section~\ref{sec:evaluation}, and broader related approaches that address
overlapping subproblems. Table~\ref{tab:mac_comparison} summarizes how
each protocol maps to the requirements identified above.

\noindent
\textbf{Tier 1: Direct Baselines (Figure~\ref{fig:MAC_protocols}).}

\textbf{\textit{ALOHA}}~\cite{abramson1970aloha,vogelgesang2021uplink}: Nodes transmit immediately upon waking without
coordination (Figure~\ref{fig:ALOHA}). No synchronization or sensing
overhead makes ALOHA simple and low-latency at sparse loads, and most
operational NTN LoRa deployments default to it for this
reason~\cite{shenoy2024cosmac}. However, collision rates grow linearly
with node density and the protocol provides no mechanism to suppress
transmissions over weak or unavailable links.

\textbf{\textit{CSMA}}~\cite{gamage2023lmac,pham2021dense,o2020practical}:
Nodes perform CAD before transmitting (Figure~\ref{fig:CSMA}),
reducing collisions effectively in terrestrial networks where nodes
are mutually audible. At satellite scale, CAD range is limited to
5--15~km~\cite{sx1276}: over 99.9\% of concurrently active nodes lie
outside each other's sensing range, making CSMA equivalent to ALOHA
over NTN footprints.

\textbf{\textit{BSMA}}~\cite{subbaraman2022bsma}: The gateway broadcasts a busy
tone upon receiving a packet, allowing distant nodes to defer regardless
of their mutual sensing range (Figure~\ref{fig:BSMA}). This solves the
large-footprint collision problem, but introduces two critical drawbacks:
(i)~continuous busy-tone transmission occupies the channel for
80--90\% of the time under high load, and (ii)~simultaneous transmit
and receive requires either a dedicated frequency channel or full-duplex
hardware, both impractical in energy-constrained NTN deployments.

\noindent
\textbf{Tier 2: Broader Related Work.}

\textbf{\textit{Slotted ALOHA and time-synchronization schemes:}}
Slotted ALOHA~\cite{polonelli2018slotted,trub2018increasing,
piyare2018demand} and NTN-adapted
variants~\cite{ma2010performance,deng2017adaptive,addaim2017enhanced}
mitigate collisions through fixed slot assignment, but require strict
timing coordination. Gateway mobility and a continuously changing node
population make maintaining synchronization across thousands of
devices impractical. A beacon-based
scheme~\cite{mojamed2024beacon} reduces some overhead but requires full LoRa packet transmissions and a dedicated downlink channel per uplink frequency, adding significant complexity without addressing link
awareness.

\textbf{\textit{RTS/CTS-based schemes:}}
RESS-IoT~\cite{ortigueira2021ress} targets satellite IoT directly,
combining beacons with RTS/CTS to reduce collisions beyond pure random
access. However, handshake overhead exceeds 200\% of payload size, and
simultaneous RTS transmissions under high node density cause
reservation-phase collisions that negate the throughput benefit.
Studies~\cite{haxhibeqiri2017lora,o2020practical,pham2021dense} address
the hidden node problem with RTS, but share the same vulnerability to
variable propagation delays and dense contention.


\vspace{-0.0035\textheight}

\textbf{\textit{NTN-specific MAC protocols:}}
CosMac~\cite{shenoy2024cosmac} coordinates uplink contention across
overlapping satellite footprints for constellation-scale throughput gains.
Spectrumize~\cite{singh2024spectrumize} exploits satellite Doppler shift
as a unique signature for packet detection and collision decoding.
SaterIoT~\cite{ren2024sateriot} improves ground-to-space links through
temporal link estimation and spatial link sharing across distributed
ground receivers. These protocols each target a specific NTN performance
dimension but assume pre-scheduled or infrastructure-assisted coordination,
making them unsuitable for the synchronization-free, single moving-gateway
scenario FSMA addresses.



As Table~\ref{tab:mac_comparison} shows, prior protocols address
collision mitigation or synchronization-free operation in isolation,
but none provides link-aware access. Protocols that do mitigate
collisions, namely BSMA, Slotted ALOHA, RTS/CTS, and CosMac/SaterIoT,
each fail on at least one of synchronization-free operation, gateway
energy efficiency, or COTS hardware compatibility. Link-aware access
is the only requirement with no checkmark across all prior protocols:
it is the column FSMA uniquely fills.

\vspace{-0.015\textheight}
\subsection{LoRa Capture Effect}
\label{sec:capture_effect}

The LoRa capture effect is a physical layer property of the SX127x
chipset~\cite{sx1276} that allows a gateway to decode a stronger packet
even when a weaker one arrives simultaneously, provided the SNR
difference between the two signals exceeds 1~dB~\cite{bor2016lora}.
Decoding begins when the receiver locks onto a preamble within a
4-symbol detection window; once locked, it commits to that signal and
ignores all subsequent arrivals~\cite{rahmadhani2018lorawan}. This
mechanism is a physical layer property available to all LoRa
protocols, not unique to FSMA. However, its reliable exploitation
requires the arrival delay spread between colliding packets to fall
within this 4-symbol locking window. In ALOHA and CSMA, two packets
can arrive at any offset up to the full packet duration (up to
404~symbols), far exceeding the locking window. So, unless the packets coming after the locking window are weaker than the locked packet, all colliding packets may be lost, making reliable capture unlikely after the locking window. FSMA bounds arrival delay spread by design, ensuring that colliding packets arrive within the capture window and enabling consistent decoding of the strongest packet. The mechanism and
performance implications are detailed in Section~\ref{sec:design}.

%% file: 3_design.tex
\section{Design of \algoname}
\label{sec:design}


\begin{figure*}[!t]
    \centering
    \begin{minipage}[t]{0.67\textwidth}
    \centering
     \begin{subfigure}[t]{0.325\textwidth}
         \centering
         \includegraphics[width=\textwidth]{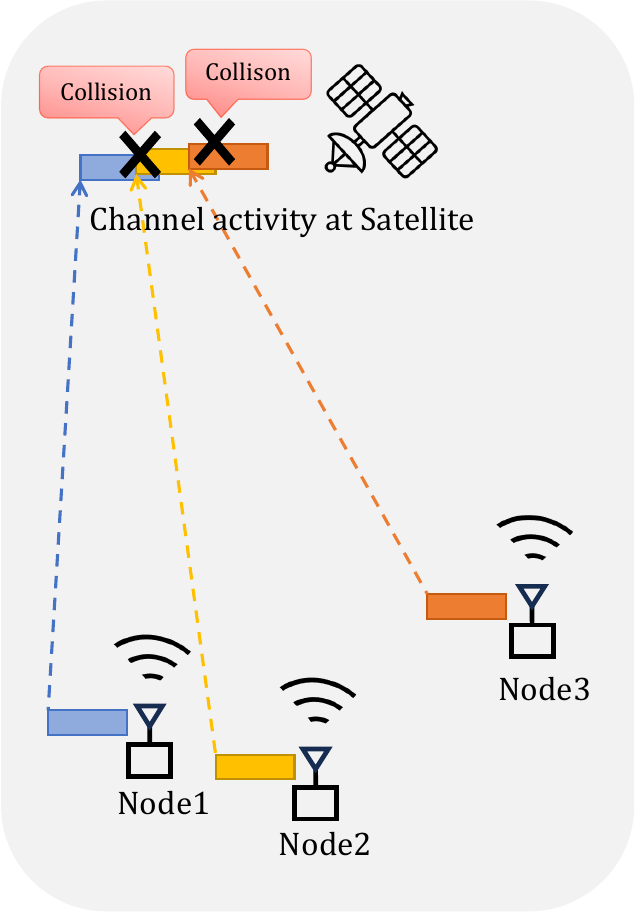}
         \caption{ALOHA}
         \label{fig:ALOHA}
     \end{subfigure}
     \hfill
     \begin{subfigure}[t]{0.325\textwidth}
         \centering
         \includegraphics[width=\textwidth]{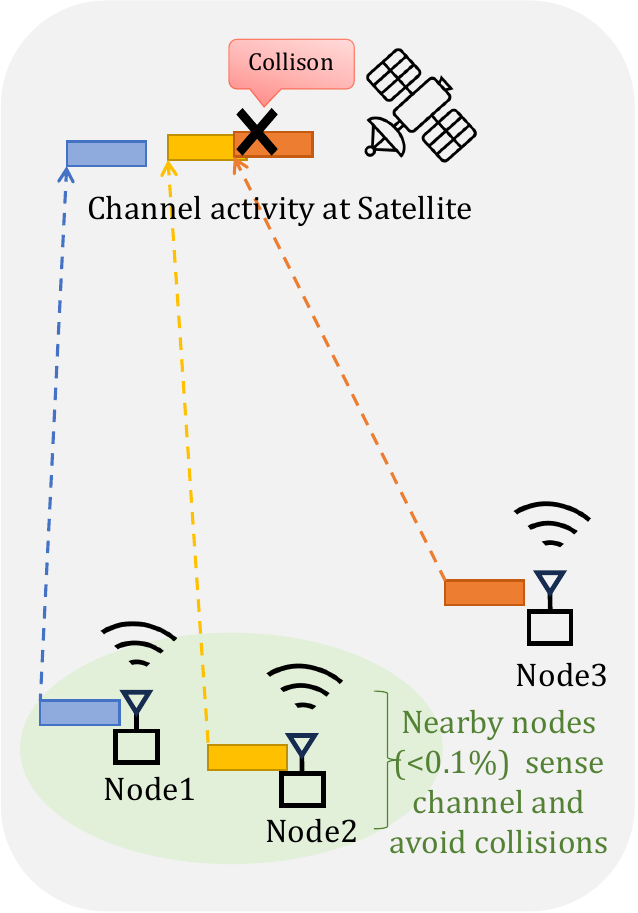}
         \caption{CSMA}
         \label{fig:CSMA}
     \end{subfigure}
     \hfill
     \begin{subfigure}[t]{0.325\textwidth}
         \centering
         \includegraphics[width=\textwidth]{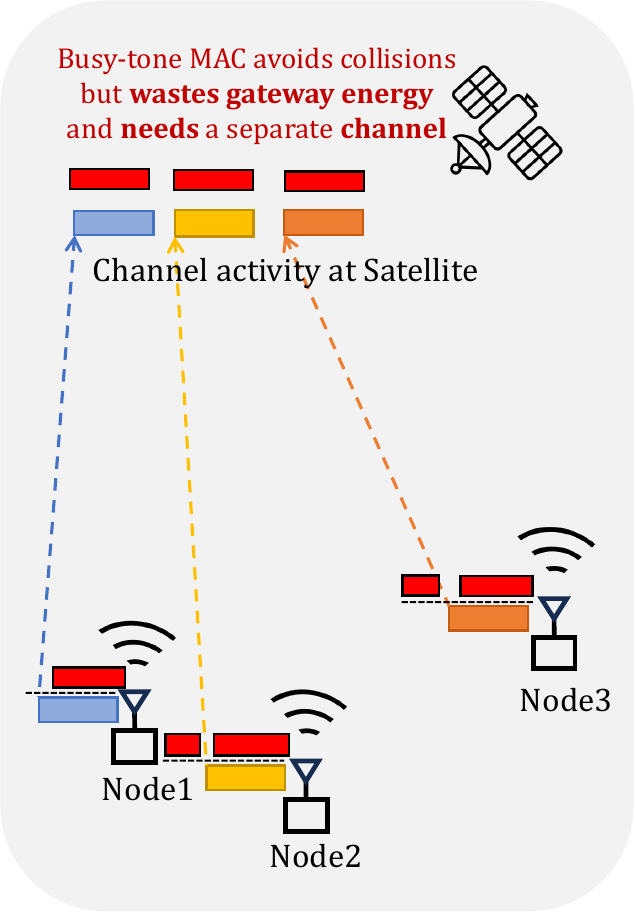}
         \caption{BSMA}
         \label{fig:BSMA}
     \end{subfigure}
     \vspace{0.01\textwidth}
     \caption{Random access MAC protocols for LoRa NTN.
     (a)~ALOHA results in frequent collisions under high density;
     (b)~CSMA avoids collisions only for nearby nodes within CAD range;
     (c)~BSMA avoids collisions but wastes gateway energy and requires
     a separate channel.}
     \label{fig:MAC_protocols}
    \end{minipage}
    \hfill
    \begin{minipage}[t]{0.32\textwidth}
    \centering
    \includegraphics[height=\textwidth]{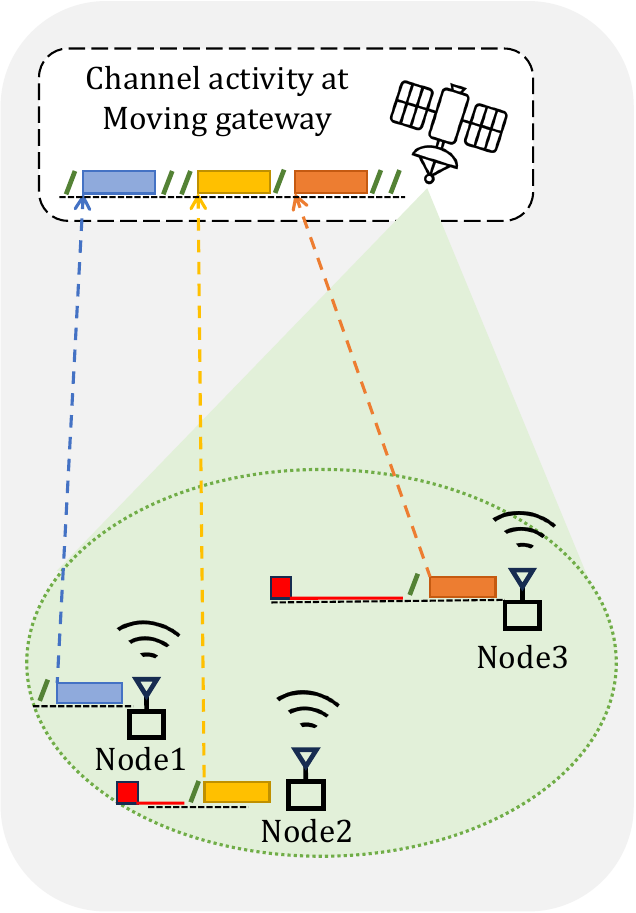}
    \caption{Proposed \algoname: a single upchirp (\textit{FreeChirp},
    shown in green) from the gateway signals channel availability and
    enables link-aware, collision-reduced access without synchronization
    or dedicated control channels.}
    \label{fig:FSMA}
    \end{minipage}
\end{figure*}

\subsection{\algoname: Free Signal Multiple Access}
\label{sec:fsma_overview}

The core challenge in NTN LoRa MAC is that any coordination signal
must originate from the gateway, since nodes cannot sense each other,
yet the gateway must coordinate access without consuming the channel
it is trying to free, without synchronization, and without hardware
changes to existing devices. At the core of FSMA is a single LoRa
upchirp called \shortname, transmitted by the gateway
whenever the channel is idle. 
FSMA operates as follows:

\begin{itemize}

\item \textbf{At the gateway (Figure~\ref{fig:gateway_process}):}
The gateway monitors the channel. If idle, it transmits a
\textit{FreeChirp}, waits for a short interval, and repeats. If
busy, it waits for a longer interval before repeating.

\item \textbf{At the nodes (Figure~\ref{fig:node_process}):}
When a node has data to send, it senses the channel for a
\textit{FreeChirp}. Upon detection, it transmits; otherwise, it
applies random backoff and retries.

\end{itemize}

As illustrated in Figure~\ref{fig:FSMA}, Node-1 detects the
\textit{FreeChirp} and transmits. Node-2 wakes during Node-1's
transmission, misses the \textit{FreeChirp}, backs off, and transmits
on the next available \textit{FreeChirp}. Node-3 similarly backs off
and transmits on a subsequent one. 

The following subsections address
three design questions that directly determine FSMA's collision
performance and energy overhead: why a single chirp; how the gateway
detects channel occupancy without additional sensing; and how nodes
reliably distinguish a \textit{FreeChirp} from a nearby node
transmission.

\subsubsection{\textbf{At Gateway: \textit{FreeChirp} Transmission
and Sensing~}}
\label{sec:gateway_design}

The gateway must solve two problems: transmit a free signal with
minimal energy and collision risk, and detect ongoing node
transmissions without additional sensing overhead.

\textbf{Why a single chirp?}
A naive approach transmits a continuous free signal until the channel
becomes busy. This has two drawbacks: a longer signal increases the
probability that multiple nodes detect it simultaneously and collide;
and it consumes more gateway energy, which is critical for
resource-constrained satellite and drone gateways. The minimum
viable free signal is one that nodes can detect reliably. Since LoRa
CAD detects a single upchirp by correlating it against a known
template~\cite{semtech_lora_cad}, one chirp is sufficient. FSMA
transmits exactly one upchirp (\shortname) upon detecting an idle channel,
minimizing both the collision window and gateway energy, then
repeats the process.

\textbf{Gateway sensing without overhead.}
Before transmitting a new \textit{FreeChirp}, the gateway must
confirm no node is still transmitting from the previous one.
Monitoring received energy levels fails at low SNRs, where most
NTN signals arrive below the noise floor. Continuous CAD at the
gateway is effective but introduces significant energy
overhead~\cite{gamage2023lmac}. Instead, FSMA decouples the
receiver and \textit{FreeChirp} transmitter into two independent
modules. The receiver module leverages the LoRa chipset's inherent
ability to output packet reception status: polling a dedicated
register exposes whether the receiver is currently locked onto a
packet or free, without any active sensing. The transmitter module polls
this status at negligible energy cost and sends the next
\textit{FreeChirp} only when the channel is confirmed free. The
specific hardware realization is described in
Section~\ref{sec:design_params}.

\begin{figure*}[!t]
  \begin{minipage}[t]{0.32\textwidth}
    \vspace{-0.18\textheight}
    \begin{subfigure}{\linewidth}
      \centering
      \includegraphics[height=0.09\textheight]{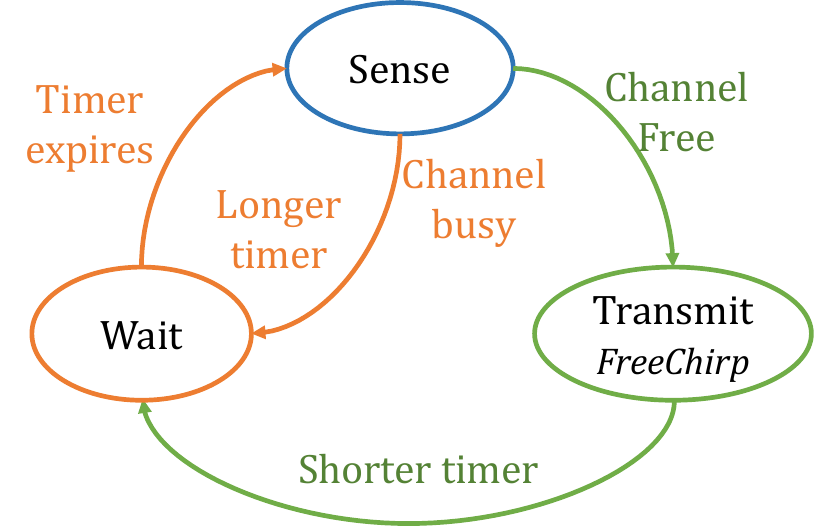}
      \caption{Process at gateway}
      \label{fig:gateway_process}
    \end{subfigure}
    \begin{subfigure}{\linewidth}
     \centering
      \includegraphics[height=0.09\textheight]{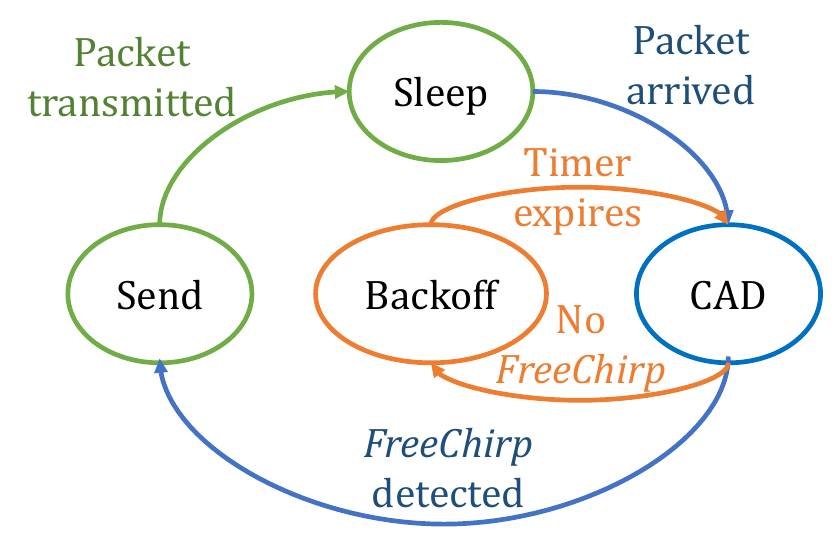}
      \caption{Process at nodes}
      \label{fig:node_process}
    \end{subfigure}
    \caption{
    Illustration of the \algoname process at the gateway
    and nodes.}
    \label{fig:fsma_full_process}
  \end{minipage}
  \hspace{0.01\textwidth}
  \begin{minipage}[t]{0.65\textwidth}
    \centering
    \begin{subfigure}[t]{0.49\textwidth}
      \includegraphics[width=\linewidth]{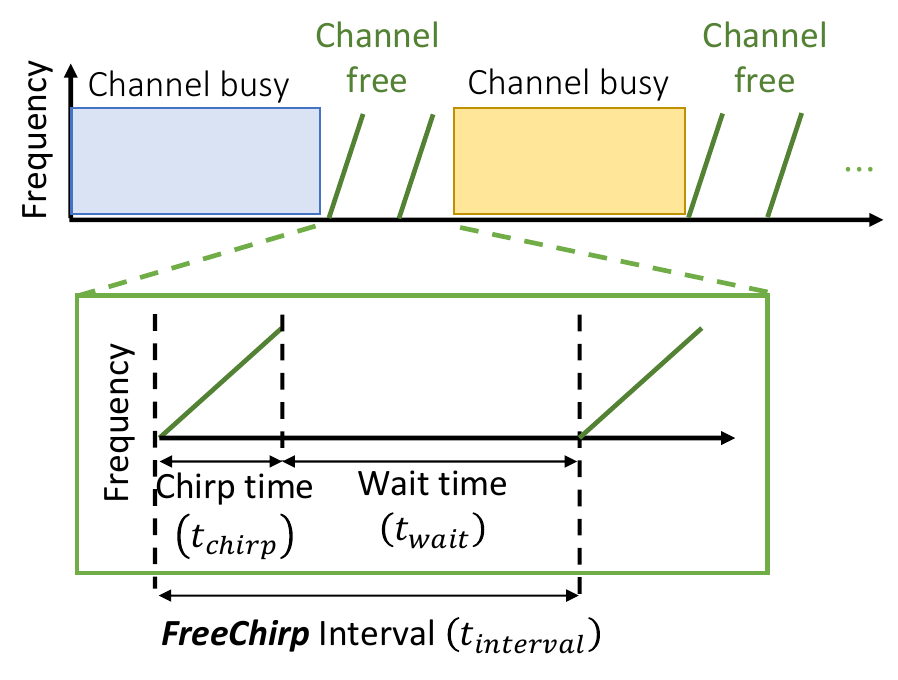}
      \caption{\textit{FreeChirp} structure at gateway}
      \label{fig:gateway_sensing}
    \end{subfigure}
    \hfill
    \begin{subfigure}[t]{0.49\textwidth}
      \includegraphics[width=\linewidth]{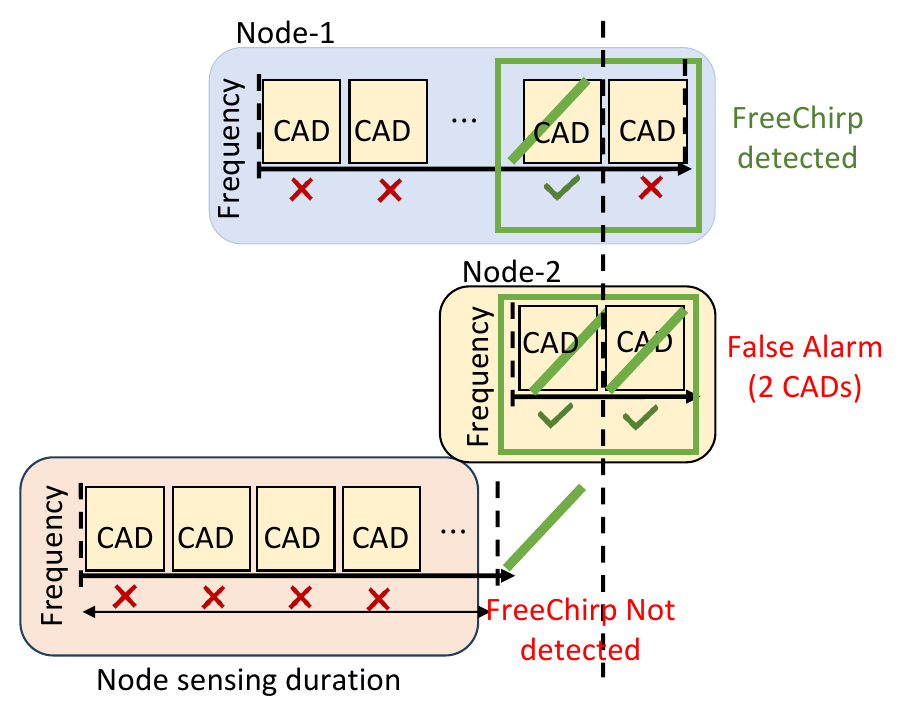}
      \caption{\textit{FreeChirp} detection at nodes}
      \label{fig:node_sensing}
    \end{subfigure}
    \caption{Overview of \algoname. (a)~At the gateway, when the
    channel is free, it periodically transmits a \textit{FreeChirp}
    and stops when the channel is busy. (b)~At the nodes, CAD is
    repeated until a positive detection is followed by a negative one,
    confirming the \textit{FreeChirp}. Nodes that do not detect within
    $t_{\textit{sense-node}}$ apply backoff. Together, this enables
    lightweight channel coordination between gateway and nodes.}
    \label{fig:freechirp_sensing}
  \end{minipage}
\end{figure*}

\textbf{\textit{FreeChirp} interval.}
After transmitting a \textit{FreeChirp}, the gateway must wait long
enough for: (i)~the chirp to propagate to nodes, (ii)~nodes to
detect it and begin transmitting, and (iii)~the return packet to
reach the gateway for detection, i.e., \textit{FreeChirp} interval is:
\begin{equation}
    T_{\textit{interval}} = t_{\textit{chirp}} + t_{\textit{wait}},
    \qquad t_{\textit{wait}} \geq 2\tau_{gn} + t_{\textit{detect}}
    \label{eq:interval}
\end{equation}
where $\tau_{gn}$ is the one-way gateway-to-node propagation delay
and $t_{\textit{detect}}$ is the preamble detection time at the
gateway receiver. The value of $t_{\textit{wait}}$ is a deployment
parameter: it must satisfy the constraint in~\eqref{eq:interval}
for the target propagation budget and receiver detection
characteristics. Section~\ref{sec:design_params} gives specific
values for our hardware and simulation configurations.
As shown in Figure~\ref{fig:gateway_sensing}, after transmitting the
\textit{FreeChirp}, the gateway waits $t_{\textit{wait}}$ while
polling the channel-occupancy signal. If the channel remains idle,
the process repeats. If a transmission is detected, the gateway
enters a longer backoff of $4 \times t_{\textit{wait}}$ to allow the
ongoing packet to complete.

\subsubsection{\textbf{At Nodes: Reliable \textit{FreeChirp}
Detection}}
\label{sec:node_design}

~Nodes transmit only upon detecting a \textit{FreeChirp} and must
distinguish it from nearby node transmissions: a false detection
causes an unnecessary transmission and a collision at the gateway.
This must work at SNRs below $-$5~dB
(Figure~\ref{fig:intro_tinygs_link_snr_analysis}), ruling out simple
energy detection. A single CAD also fails: ongoing node
transmissions consist of continuous upchirps, producing the same
positive CAD result as a \textit{FreeChirp}.
The key insight is that a \textit{FreeChirp} and a node transmission
have a structurally different temporal signature. A \textit{FreeChirp}
is a single chirp followed by a long silent interval; a node
transmission is a continuous stream of upchirps. Two consecutive CADs
therefore produce a unique pattern for a \textit{FreeChirp}:
positive-then-negative. For a node transmission:
positive-then-positive. FSMA uses this two-stage verification to
confirm a \textit{FreeChirp} before transmitting.

\textbf{Stage 1 -- Detection:}
The node continuously performs CAD until a positive detection or the
sensing window expires. Since the gateway is guaranteed to transmit a
\textit{FreeChirp} within every $T_{\textit{interval}}$ when the
channel is idle, the sensing window is set to:
\begin{equation}
    T_{n\textit{Sense}} = T_{\textit{interval}}
    \label{eq:nSense}
\end{equation}
This ensures no \textit{FreeChirp} is missed during any free channel
period.

\textbf{Stage 2 -- Verification:}
Upon a positive detection, the node immediately performs a second
CAD. A negative result confirms a \textit{FreeChirp}; a positive
result indicates an ongoing node transmission. As shown in
Figure~\ref{fig:node_sensing}, Node-1 observes
positive-then-negative and transmits. Node-2 observes two consecutive
positives, correctly identifies an ongoing transmission, and defers.
Node-3's sensing window expires without a positive detection,
triggering backoff.

\subsection{\algoname Reduces Collisions and Packet Loss}
\label{sec:collision_reduction}


\subsubsection{\textbf{Reduced Collision Window~}}
\label{sec:collision_window}

~In baseline protocols, any node that wakes during an ongoing
reception and transmits causes a collision. A node that started
transmitting up to $T_P$ before the current packet creates a
tail-overlap collision; a node that starts up to $T_P$ after creates
a head-overlap collision. The collision window therefore spans twice
the packet duration:
$    T_{cw}^{\textit{ALOHA}} = 2T_P  $
where $T_P$ spans 20.25 to 404.25~symbols for 0 to 192-byte payloads
at SF10~\cite{wiki_swarm,lora_calculator}
(Figure~\ref{fig:baseline_collision_space}).
FSMA transmits a new \textit{FreeChirp} only after confirming no
transmission was triggered by the previous one. Simultaneous
transmissions are therefore restricted to nodes that detect the
\emph{same} \textit{FreeChirp}, reducing the collision window from
$2T_P$ to the node sensing period
(Figure~\ref{fig:fsma_collision_space}):
\begin{equation}
    T_{cw}^{\textit{FSMA}} = T_{n\textit{Sense}} =
    t_{\textit{chirp}} + 2\tau_{gn} + t_{\textit{detect}}
    \label{eq:fsma_cw}
\end{equation}
FSMA \emph{decouples} the collision window from packet length: $T_{cw}$
no longer scales with payload size but depends only on propagation
geometry and receiver detection latency, both of which are small
fractions of $T_P$ across all LoRa deployments. 
Typically, $T_{n\textit{Sense}}$ is between 4--6 node symbols, resulting in a 6--200$\times$ collision window reduction depending on the payload size. 
For residual contention when multiple nodes detect the same
\textit{FreeChirp}, FSMA integrates progressive
backoff~\cite{wiki_expo_backoff,rohm2009dynamic,vemasaniexponential}
to further spread contention within the sensing window.

\subsubsection{\textbf{Analytical Throughput Bound}}
\label{sec:analysis}

~With Poisson arrivals and offered load $G = N\lambda T_P$, a tagged
packet succeeds only if no competing node transmits during its
vulnerability window $T_{cw}$. This gives normalized throughput:
\begin{equation}
    S(G) = G \cdot e^{-G \cdot T_{cw}/T_P}
    \label{eq:throughput}
\end{equation}
This is a conservative lower bound, treating every collision as
unrecoverable. The key insight is that every protocol reduces to
Equation~\eqref{eq:throughput} with a different $T_{cw}$: shrinking
$T_{cw}$ directly improves throughput (Table~\ref{tab:tcw}).

\smallskip
\noindent\textbf{Baselines: $T_{cw}$ scales with $T_P$.}
In ALOHA, $T_{cw} = 2T_P$ as derived above. CSMA sensing eliminates
backward exposure, collapsing $T_{cw}$ to $\tau_{nn}$, the
inter-node propagation delay, which is highly effective
terrestrially. However, when nodes are farther apart than the CAD
range, sensing fails. In NTN, the probability of hidden nodes
($P_H$) approaches unity: CAD range is 3--15~km while a satellite
footprint spans thousands of kilometers, giving $P_H > 0.99$.
Substituting into
$T_{cw} = P_H \cdot 2T_P + (1-P_H) \cdot \tau_{nn}$
recovers $T_{cw} \approx 2T_P$. \emph{CSMA analytically degrades to
ALOHA in NTN regardless of protocol parameters.}

\smallskip
\noindent\textbf{FSMA: $T_{cw}$ decoupled from $T_P$.}
Only nodes detecting the same \textit{FreeChirp} can collide, so
$T_{cw}$ is no longer the packet duration but the node sensing
period: $T_{cw}$ = $T_{n\textit{Sense}}$, a small value independent
of payload size. Section~\ref{sec:design_params} gives the numerical
instantiation. Hardware measurements show 2--3$\times$ throughput
improvement, which is a conservative reflection of the analytical
gain. 



\begin{table}[t]
\small\centering
\setlength{\tabcolsep}{3pt}
\vspace{0.005\textheight}
\caption{Collision window $T_{cw}$ and normalized throughput $S(G)$
per protocol. $\eta = T_P/(T_P + t_{\textit{interval}})$ accounts
for \textit{FreeChirp} overhead.
$^\dagger$In NTN, $P_H > 0.99$, collapsing CSMA to ALOHA.}
\label{tab:tcw}
\begin{tabular}{lcc}
\toprule
Protocol & $T_{cw}$ & $S(G)$ \\
\midrule
ALOHA
  & $2T_P$
  & $G\,e^{-2G}$ \\[3pt]
CSMA (ideal)
  & $\tau_{nn}$
  & $G\,e^{-G\tau_{nn}/T_P}$ \\[3pt]
CSMA (NTN)$^\dagger$
  & $P_H{\cdot}2T_P {+} (1{-}P_H){\cdot}\tau_{nn}$
  & $G\,e^{-G \cdot T_{cw}/T_P}$ \\[3pt]
FSMA (ideal)
  & $2\tau_{gn}$
  & $\eta\,G\,e^{-2G\tau_{gn}/T_P}$ \\[3pt]
FSMA (practical)
  & $2\tau_{gn} {+} \delta T_{\textit{det}}$
  & $\eta\,G\,e^{-G(2\tau_{gn}+\delta T_{\textit{det}})/T_P}$ \\
\bottomrule
\end{tabular}
\vspace{-0.01\textheight}
\end{table}

\subsubsection{\textbf{Reliable Exploitation of the Capture Effect}}
\label{sec:capture_design}

~As established in Section~\ref{sec:capture_effect}, the capture
effect decodes a stronger packet during a collision only if the
arrival delay spread falls within the SX127x 4-symbol locking
window. In ALOHA and CSMA, packets can arrive at any offset up to
$2T_P$, far exceeding this window and making reliable capture
unlikely. FSMA bounds arrival delay spread by design: nodes
detecting the same \textit{FreeChirp} transmit nearly simultaneously,
so packets arrive offset only by propagation delay differences:
\vspace{-0.01\textheight}
\begin{equation}
    t_{\textit{arrival-FSMA}} = 2 \times \max(t_{pd_i} - t_{pd_j})
    \label{eq:arrival_spread}
\end{equation}
For any LoRa deployment, this difference is a small fraction of one
symbol, well within the locking window. This allows FSMA to decode the strongest packet even during 
collisions, unlike baseline approaches. The residual failure case of equal power
arrivals is a fundamental hardware limit, not specific to
FSMA~\cite{bor2016lora}.

\vspace{-0.01\textheight}
\subsection{\algoname Enables Link-Aware Transmissions}
\label{sec:link_aware}

In NTN scenarios, link quality fluctuates by more than 10~dB within
a single pass and drops to a complete outage between passes
(Figure~\ref{fig:intro_tinygs_link_snr_analysis}). Nodes transmitting
during low-SNR or outage periods waste airtime regardless of
collision avoidance. Pre-computing pass schedules addresses gateway availability
but requires per-node firmware updates as trajectories change and
cannot account for within-pass SNR fluctuations.

FSMA addresses this by transmitting \textit{FreeChirp} at a lower SF
than the node uplink and exploiting channel reciprocity: if a node
detects the gateway's lower-SF downlink chirp, the uplink channel at
the higher SF supports a successful transmission with high
probability. \textit{FreeChirp} therefore acts simultaneously as a
channel-free indicator and an implicit link quality probe, with no
additional signaling. Nodes with weak links, blockage, or outside
instantaneous coverage receive no \textit{FreeChirp} and apply
backoff, preventing futile transmissions. Transmitting at a lower SF
also reduces the \textit{FreeChirp} ground coverage, further
limiting the contending population to nodes with strong links. The
specific SF configuration used in our deployment, and its validation,
are given in Section~\ref{sec:design_params}.

A natural concern is whether limiting \textit{FreeChirp} coverage
permanently excludes distant nodes. Unlike static terrestrial
gateways, a moving satellite or drone continuously shifts its
coverage footprint. A node excluded at one gateway position gains
access as the geometry changes, encouraging transmission at higher
elevation angles when the link is strong. Long-term fairness emerges
from gateway mobility without any explicit fairness mechanism.

%% file: 4_Implementation.tex
\vspace{-0.015\textheight}
\section{Implementation}
\label{sec:implementation}

\begin{figure}[!t]
 \centering
 \begin{subfigure}[t]{0.23\textwidth}
     \centering
     \includegraphics[height=0.17\textheight]{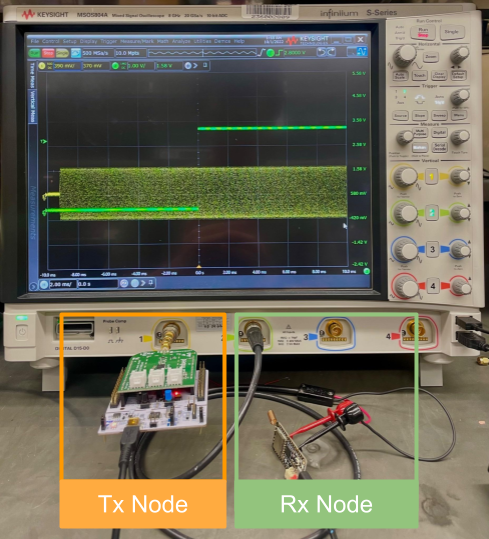}
     \caption{Oscilloscope setup: Tx to CH1, Rx IO pin to CH2.}
     \label{fig:eval_detection_setup}
 \end{subfigure}
 \hfill
 \begin{subfigure}[t]{0.23\textwidth}
     \centering
     \includegraphics[height=0.17\textheight]{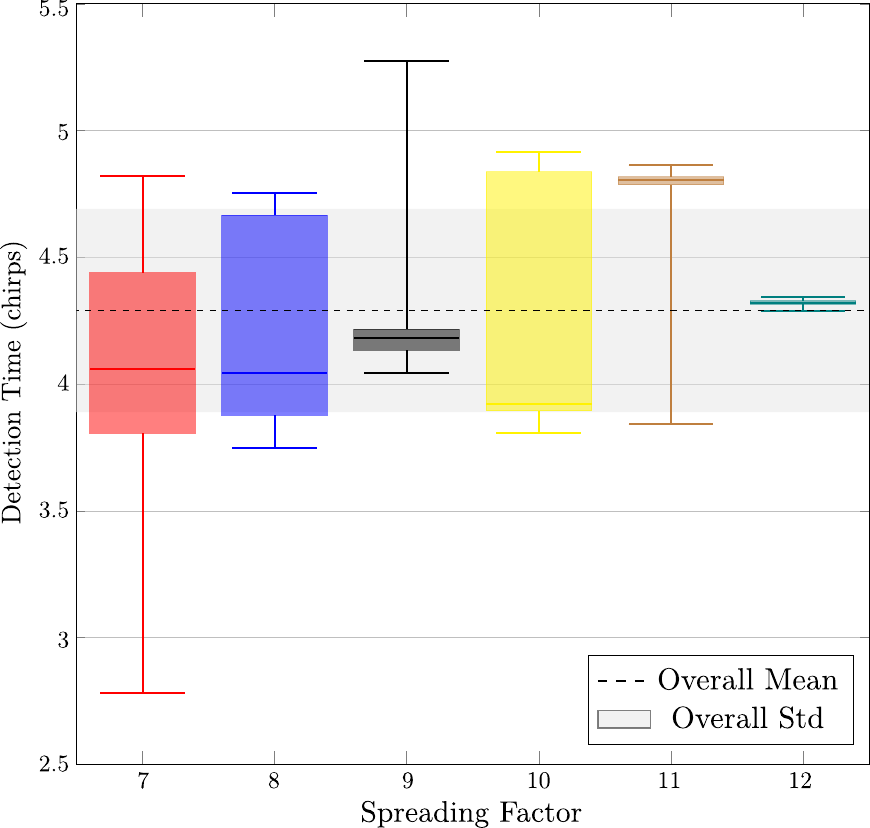}
     \caption{Detection delay in chirps across SFs.}
     \label{fig:eval_detection_plot}
 \end{subfigure}
\caption{SX1276 preamble detection characterization. Detection
occurs within 3--5 chirps across all tested SFs, validating the
$t_{\textit{wait}}$ design parameter.}
\label{fig:eval_detection}
\vspace{-0.01\textheight}
\end{figure}

\subsection{Design Parameters}
\label{sec:design_params}

Section~\ref{sec:gateway_design} defines $t_{\textit{wait}}$ as a
deployment parameter satisfying
$t_{\textit{wait}} \geq 2\tau_{gn} + t_{\textit{detect}}$. This
section instantiates that constraint for the hardware and simulation
configurations evaluated in this paper.

\noindent\textbf{Wait time:}
To measure $t_{\textit{detect}}$, we connect an Adafruit SX1276
receiver to a Keysight MSOS804A oscilloscope: CH1 monitors the
transmitter RF output and CH2 monitors the IO pin reflecting bit~0
of the \textit{RegModemStat} register~\cite{sx1276}, which is
asserted when a valid preamble is detected. Detection delay is
measured as the time from RF power onset to the IO pin rising edge.
Across 40 trials per SF, the signal-detected bit rises consistently
within 3--5 chirp symbols
(Figure~\ref{fig:eval_detection}), giving
$t_{\textit{detect}} \leq 5 \times t_{n\textit{Sym}}$. At NTN
distances, one-way propagation satisfies $\tau_{gn} <
t_{n\textit{Sym}}$, so the constraint evaluates to
$t_{\textit{wait}} \geq (1 + 5) \times t_{n\textit{Sym}}$. Hardware
experiments therefore use $t_{\textit{wait}} = 6 \times
t_{n\textit{Sym}}$, providing one symbol of guard margin.



\begin{figure}[!t]
         \centering
         \begin{subfigure}[t]{0.23\textwidth}
             \centering
             \includegraphics[width=\textwidth]{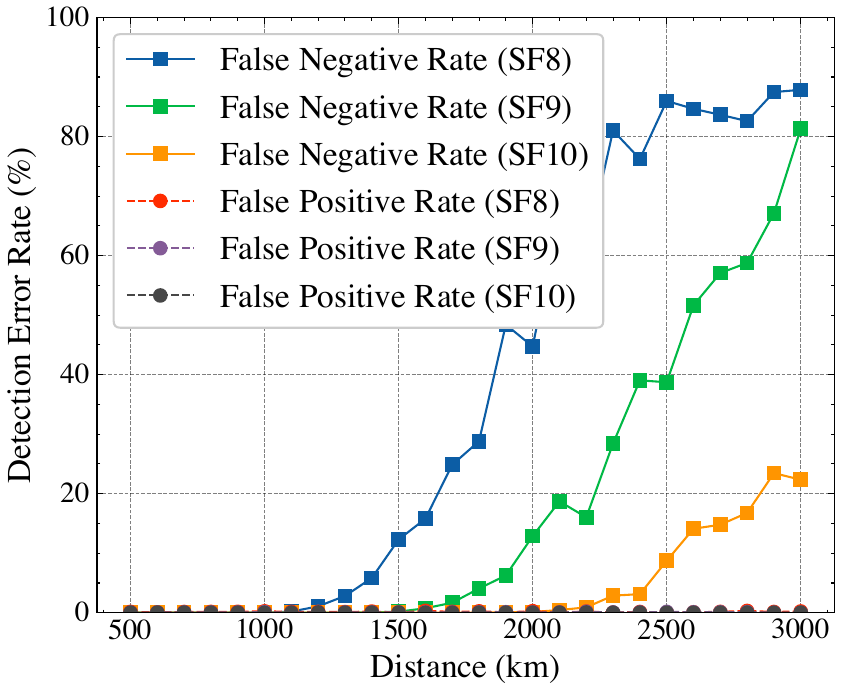}
             \caption{FreeChirp detection rate at nodes vs. distance.}
             \label{fig:eval_cad_detection_efficiency.pdf}
         \end{subfigure}
         \hfill
         \begin{subfigure}[t]{0.23\textwidth}
             \centering
             \includegraphics[width=\textwidth]{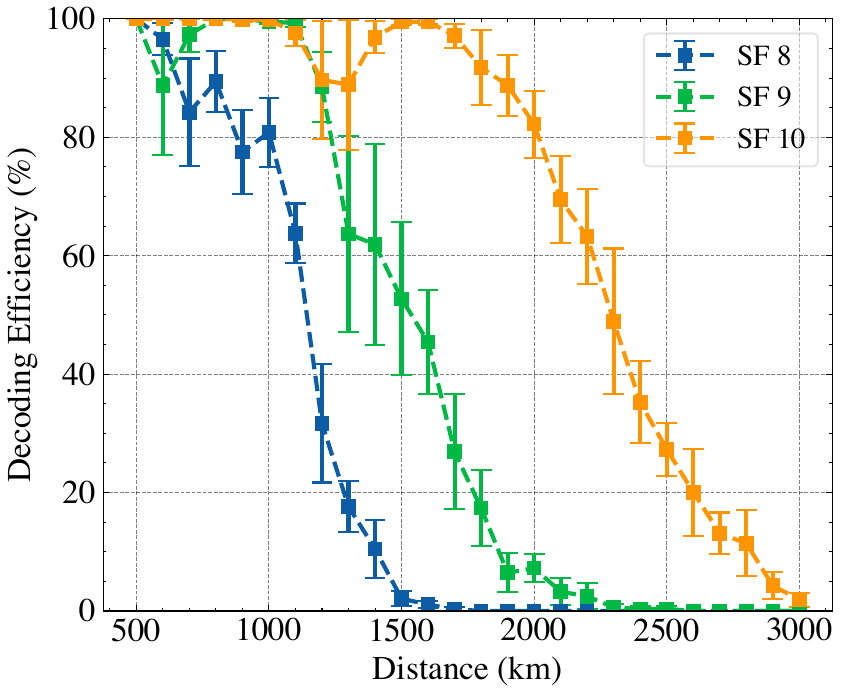}
             \caption{Decoding efficiency at gateway vs. distance.}
             \label{fig:eval_decoding_efficiency}
         \end{subfigure}
        \caption{\shortname detection at nodes and packet decoding
        efficiency at the gateway vs. distance. SF9 and SF10 provide
        reliable coverage up to 2000~km for \shortname and node uplink,
        respectively.}
        \vspace{-0.01\textheight}
        \label{fig:eval_freechirp_detection}
\end{figure}

\noindent\textbf{FreeChirp SF selection:}
The gateway transmits \textit{FreeChirp} at SF9; nodes transmit
uplink data at SF10. Figure~\ref{fig:eval_freechirp_detection} shows
that SF9 achieves zero missed detections up to 2000~km with the
lowest airtime among tested SFs, making it the optimal
\textit{FreeChirp} configuration for our NTN deployment. SF10
provides the best balance of range and capacity for node uplinks.
These results validate the SF9/SF10 configuration used throughout
Section~\ref{sec:evaluation}.

\noindent\textbf{Collision window at this configuration:}
Substituting into Equation~\eqref{eq:fsma_cw}, $T_{cw}^{\textit{FSMA}}
\approx 6 \times t_{n\textit{Sym}}$ versus $2T_P = 41$--$809$
symbols for baselines: a $6\times$--$200\times$ reduction depending
on payload size. At SF10 with a 20-byte payload
($T_P = 166$~ms, $\tau_{gn} \approx 5$~ms,
$\delta T_{\textit{det}} \approx 33$~ms), $T_{cw}/T_P$ drops from
2.0 to 0.26, a theoretical $7.7\times$ lower bound on throughput
gain over ALOHA.

\begin{figure}[!t]
    \centering
        \centering
        \begin{subfigure}[t]{0.23\textwidth}
             \centering
             \includegraphics[height=0.18\textheight]{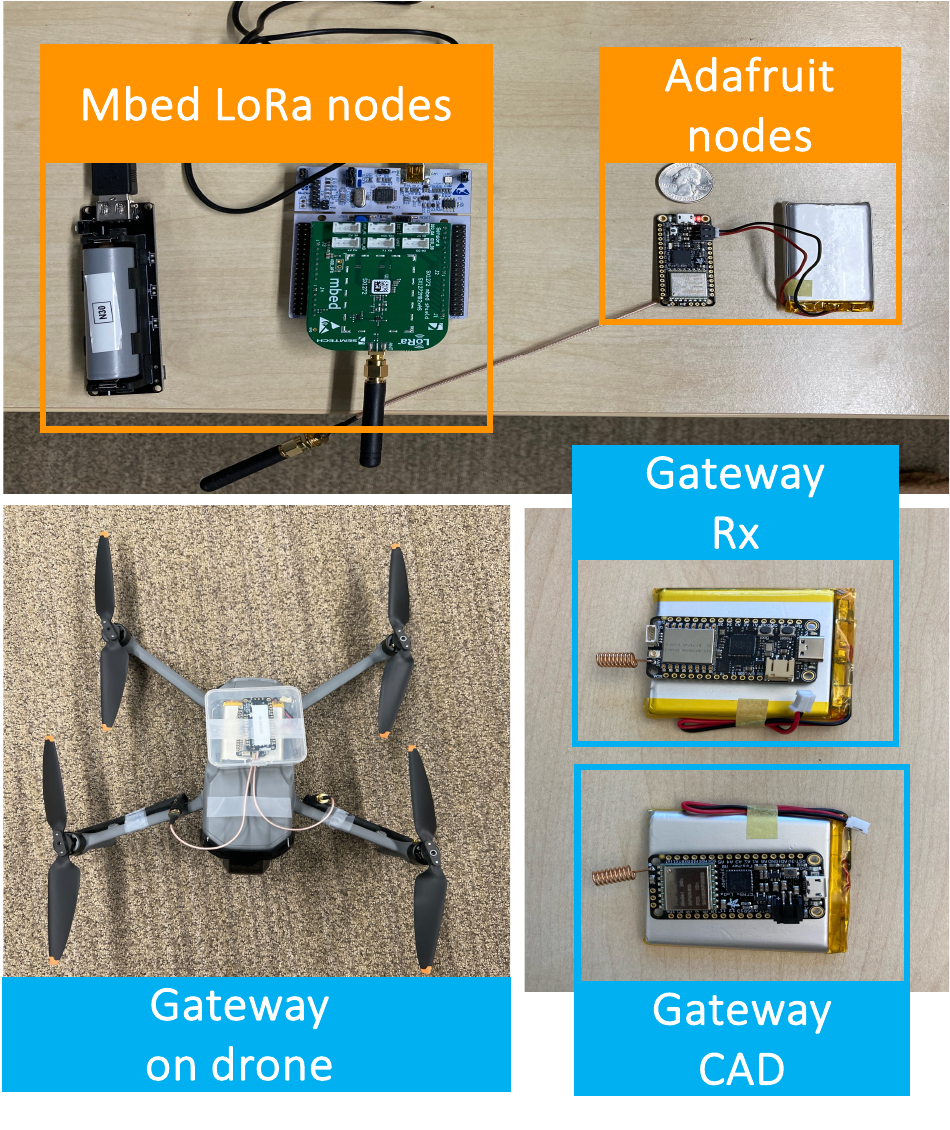}
             \caption{Hardware testbed: Mbed and Adafruit LoRa nodes.}
             \label{fig:hardware_setup}
         \end{subfigure}
         \hfill
         \begin{subfigure}[t]{0.23\textwidth}
             \centering
             \includegraphics[height=0.18\textheight]{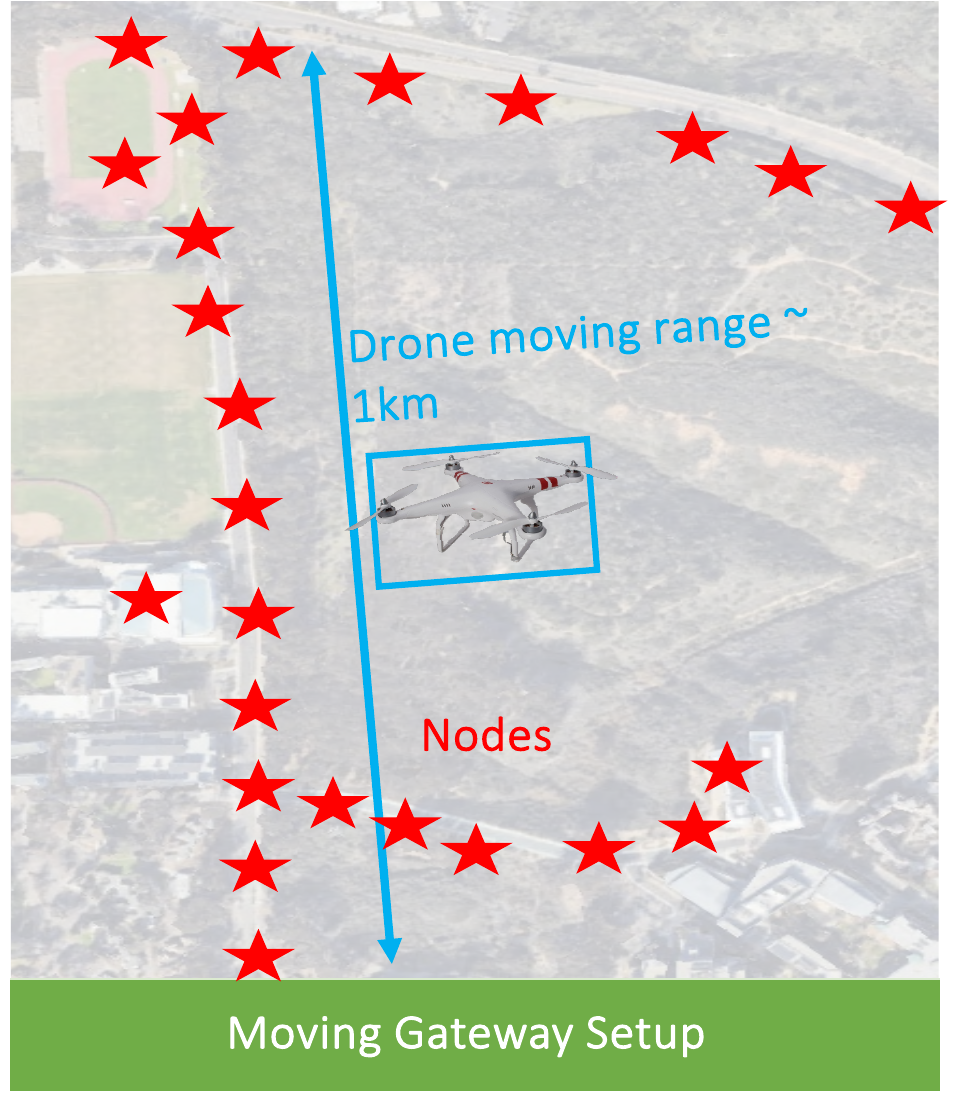}
             \caption{Moving gateway (drone): node deployment and
             flight path.}
             \label{fig:drone_moving_scenario}
         \end{subfigure}
         \caption{Hardware setup for evaluating \algoname. (a)~ Semtech (Mbed)
         and Adafruit (M0, RP2040) nodes; two Adafruit RP2040 boards form the
         gateway. (b)~Moving gateway test scenario.}
         \vspace{-0.02\textheight}
        \label{fig:hw_sim}
\end{figure}

\vspace{-0.01\textheight}
\subsection{Hardware Testbed}
\label{sec:hardware_testbed}

\noindent\textbf{Nodes:}
Two types of COTS LoRa devices serve as nodes: Adafruit Feather M0
boards with SX1276 modules and Mbed STM32
boards with SX1272 modules, both battery-powered
(Figure~\ref{fig:hardware_setup}). All nodes are configured with SF10,
125~kHz bandwidth, 20-byte payload, CR4/8, explicit header, and CRC
enabled, meeting the sensitivity requirements of NTN IoT links at
extremely low SNRs. Packet arrivals follow a Poisson process governed
by a per-node duty-cycle constraint. Both device families natively
support CAD~\cite{semtech_lora_cad,gamage2023lmac,subbaraman2022bsma}.
Enabling FSMA requires only a firmware update that reverses the CAD
logic: a node must observe a positive CAD followed by a negative CAD
to confirm a \textit{FreeChirp} (Figure~\ref{fig:node_sensing}).

\noindent\textbf{Gateway:}
The gateway consists of two off-the-shelf Adafruit Feather RP2040
boards. One board receives node packets and
stores up to 8~MB of experimental data in onboard memory for remote
retrieval. Channel occupancy is determined without additional sensing
by monitoring bit~0 of the SX127x \textit{RegModemStat}
register~\cite{sx1276} through an external trigger pin, which is
asserted when a preamble is detected and released upon reception
completion. The second board monitors this trigger and transmits a
\textit{FreeChirp} periodically when the pin indicates a free channel.

\begin{figure}[!t]
    \centering
    \begin{subfigure}[t]{0.23\textwidth}
         \centering
         \includegraphics[height=0.18\textheight]{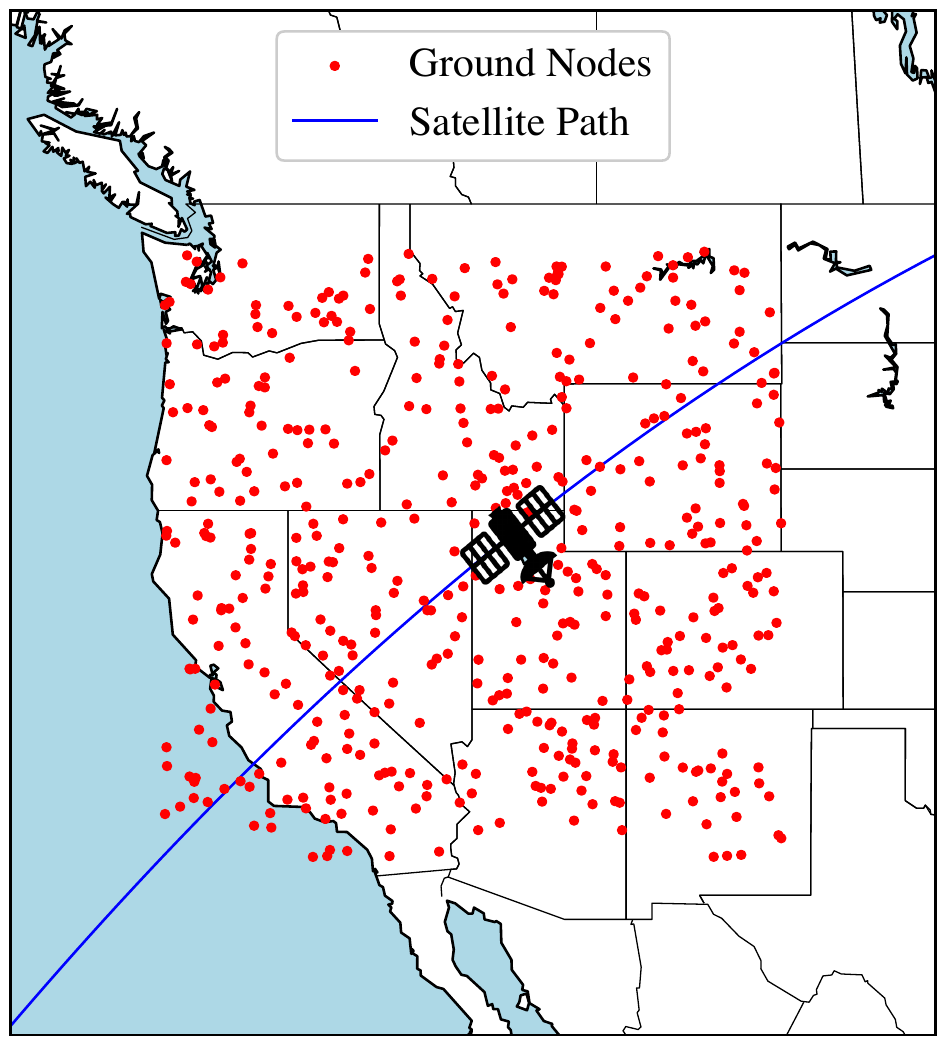}
         \caption{Node deployment and satellite path (TLE-based).}
         \label{fig:sim_gateway_nodes_visual}
     \end{subfigure}
     \hfill
     \begin{subfigure}[t]{0.23\textwidth}
         \centering
         \includegraphics[height=0.18\textheight]{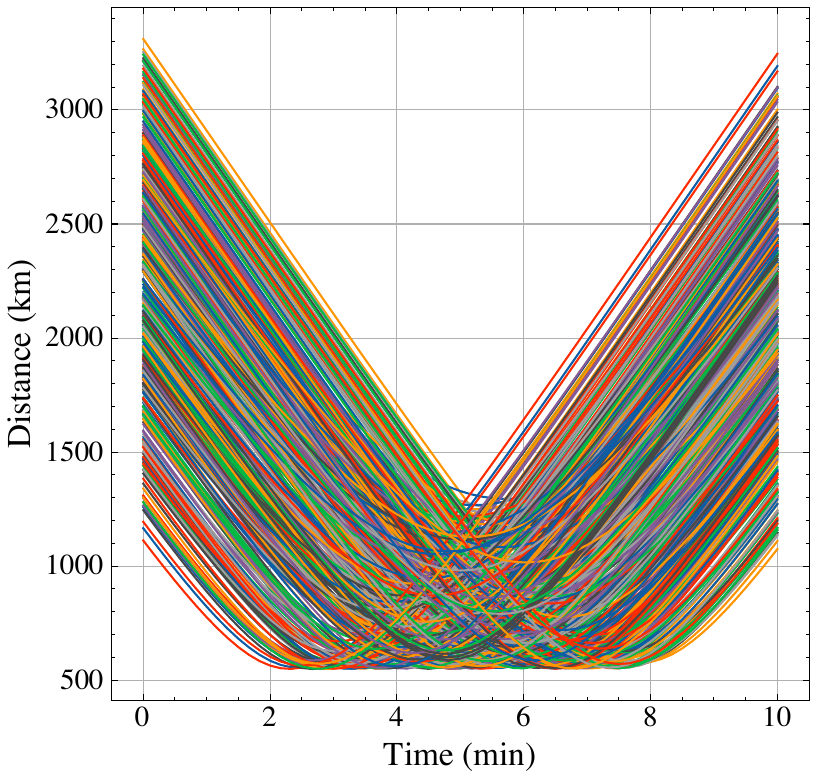}
         \caption{Satellite-to-node distance over a 10-minute
         pass.}
         \label{fig:sim_gateway_nodes_distances_plot}
     \end{subfigure}
     \caption{\textit{NTNLoRa} simulator: nodes distributed across a
     geographic region with a TLE-based moving gateway.}
     \vspace{-0.02\textheight}
    \label{fig:python_simulator}
\end{figure}

\begin{table}[t]
\small\centering
\setlength{\tabcolsep}{4pt}
\caption{Experiment and simulation parameters}
\label{tab:params}
\begin{tabular}{ll}
\toprule
\textbf{Category} & \textbf{Parameter (Value)} \\
\midrule
Scenario   & Duration - 600~sec, Frequency - 920~MHz\\
           & Nodes - 25 (HW) / variable (Sim) \\
           & Duty cycle - variable (HW) / 0.1\%  (Sim) \\
\midrule
LoRa PHY   & SF10, BW 125~kHz, CR 4/8, 20~Bytes payload \\
           & \textit{FreeChirp} SF: 9 \\
\midrule
Backoff    & Exponential; init: packet length; \\
           & reset after $>$100$\times$ initial window \\
\bottomrule
\end{tabular}
\vspace{-0.01\textheight}
\end{table}

\noindent\textbf{Single-chirp transmission on COTS hardware:}
COTS LoRa devices cannot transmit a single chirp by default: even a
zero-payload packet requires multiple chirps for the preamble, header,
and CRC. FSMA enables single-chirp transmission by using OS interrupt
timers to halt the radio after exactly one chirp duration, requiring
only a software change with no hardware modification.

\noindent\textbf{Controller:}
An additional LoRa device controls the entire testbed remotely,
configuring per-node parameters at runtime including SF, BW, CR,
center frequency, transmit power, MAC protocol, and traffic type. A
trigger message starts the experiment; nodes run for the configured
duration and report per-node statistics including transmitted packet
count, wait time, and energy metrics for post-processing. This testbed
extends the base version from~\cite{subbaraman2022bsma}.

\noindent\textbf{Moving gateway setup:}
To replicate NTN link variability and dynamic coverage, the gateway
is mounted on a drone and 25 nodes are deployed across a campus-scale
area spanning over 1~km (Figure~\ref{fig:hardware_setup}). The drone
completes a full loop in 4~minutes at 10~m/s
(Figure~\ref{fig:drone_moving_scenario}), producing continuously
varying gateway-to-node distances and link conditions representative
of real satellite and drone IoT deployments.

\noindent\textbf{Static gateway setup:}
A static setup isolates FSMA's collision mitigation from link
variability. Sixteen nodes and a static gateway are used, with
attenuators at transmitter and receiver to emulate low-SNR links.
Since COTS gateways decode one packet at a time, any simultaneous
transmission causes a collision. To emulate large-scale NTN
contention without requiring thousands of physical nodes, the offered
load per node is swept up to 30\% (up to 500\% total offered network
load), reproducing the collision conditions that arise under strict
duty-cycle constraints across large deployments.

\vspace{-0.01\textheight}
\subsection{\textit{NTNLoRa}: Python PHY-MAC Simulator}
\label{sec:simulator}

To validate FSMA at scales beyond the 25-node hardware testbed, we
develop \textit{NTNLoRa}, a custom Python-based PHY-MAC simulator for satellite
and drone gateway IoT scenarios. \textit{NTNLoRa} models end-to-end behavior
including packet arrivals, LoRa waveform parameters, satellite channel
dynamics, CAD detection, packet reception, and the LoRa capture
effect, closely matching observed hardware behavior.
Nodes are distributed across user-defined geographic regions specified
by latitude and longitude coordinates
(Figure~\ref{fig:sim_gateway_nodes_visual}). Satellite movement
follows TLE-based orbital mechanics, reproducing realistic
trajectories and link dynamics.
Figure~\ref{fig:sim_gateway_nodes_distances_plot} shows
satellite-to-node distance variation over a 10-minute pass, with each
line representing a different node. \textit{NTNLoRa} evaluates system-level
metrics including throughput, PRR, energy efficiency, and channel
utilization, as well as physical-layer behavior including CAD
detection accuracy and collision decoding outcomes. The simulator and
complete FSMA firmware are open-sourced
at~\href{https://github.com/ucsdwcsng/fsma-lora}{\underline{github.com/ucsdwcsng/fsma-lora}}.

%% file: 5_evaluation.tex
\vspace{-0.01\textheight}
\section{Evaluation}
\label{sec:evaluation}

\noindent\textbf{System parameters.}
Table~\ref{tab:params} summarizes the configuration used across
hardware experiments and simulations, grouped by scenario, LoRa PHY,
and backoff parameters.



\noindent
\textbf{Baselines.}
Simulations compare FSMA against ALOHA, BSMA, and CSMA variants with
hearing ranges of 10~km (realistic), 500~km, 1000~km, and 2000~km.
The 2000~km variant is an oracle: unreachable in practice but useful
as a performance upper bound. Propagation delay is modeled in all
variants (e.g., a 1500~km transmission is heard after $\approx$5~ms).
Under NTN mobility, CSMA at realistic hearing range degrades to ALOHA
due to the hidden node problem. BSMA improves collision avoidance but
incurs prohibitive gateway energy overhead from continuous busy-tone
transmission. Hardware experiments use ALOHA as the sole baseline,
equivalent to CSMA with $>$99\% hidden nodes at satellite scale.

\begin{figure}[!t]
     \centering
     \begin{subfigure}[t]{0.23\textwidth}
         \centering
         \includegraphics[height=0.15\textheight]{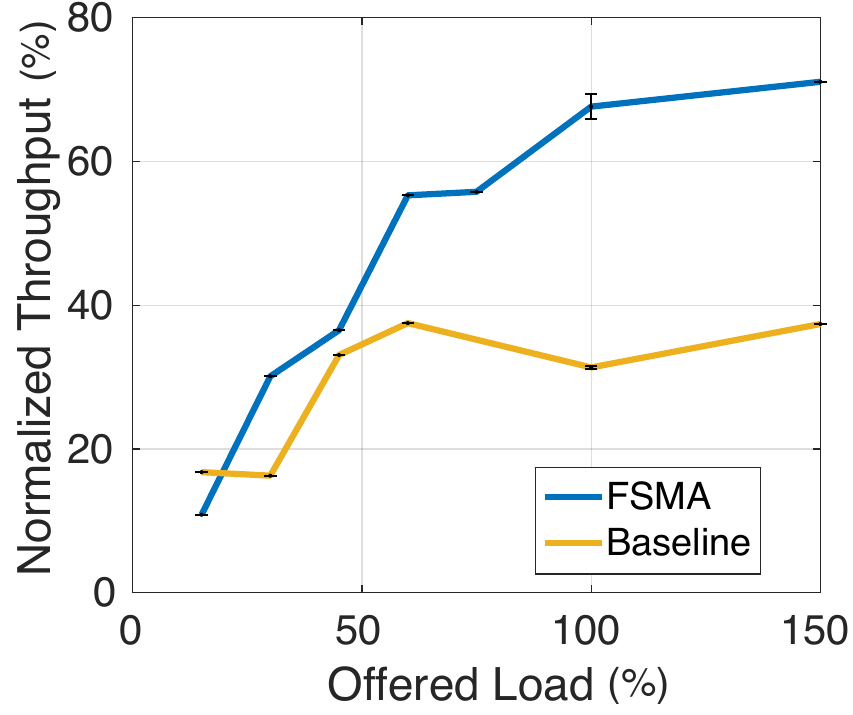}
         \caption{Throughput}
         \label{fig:eval_throughput}
     \end{subfigure}
     \hfill
     \begin{subfigure}[t]{0.23\textwidth}
         \centering
         \includegraphics[height=0.15\textheight]{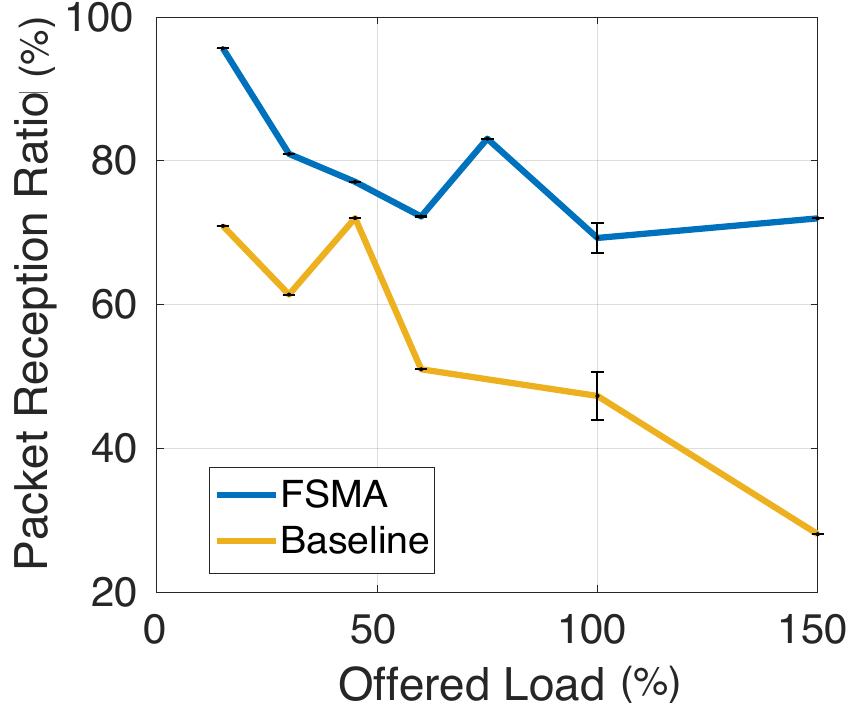}
         \caption{Packet reception ratio}
         \label{fig:eval_PRR}
     \end{subfigure}
     \setlength{\belowcaptionskip}{-4pt}
     \caption{Hardware experiments: moving gateway (drone).}
      \vspace{-0.01\textheight}
     \label{fig:eval_hardware_drone}
\end{figure}

\subsection{Hardware Experiments: Static and Moving}
\label{sec:hw_eval}

\begin{figure}[t]
     \centering
     \begin{subfigure}[t]{0.23\textwidth}
         \centering
         \includegraphics[height=0.15\textheight]{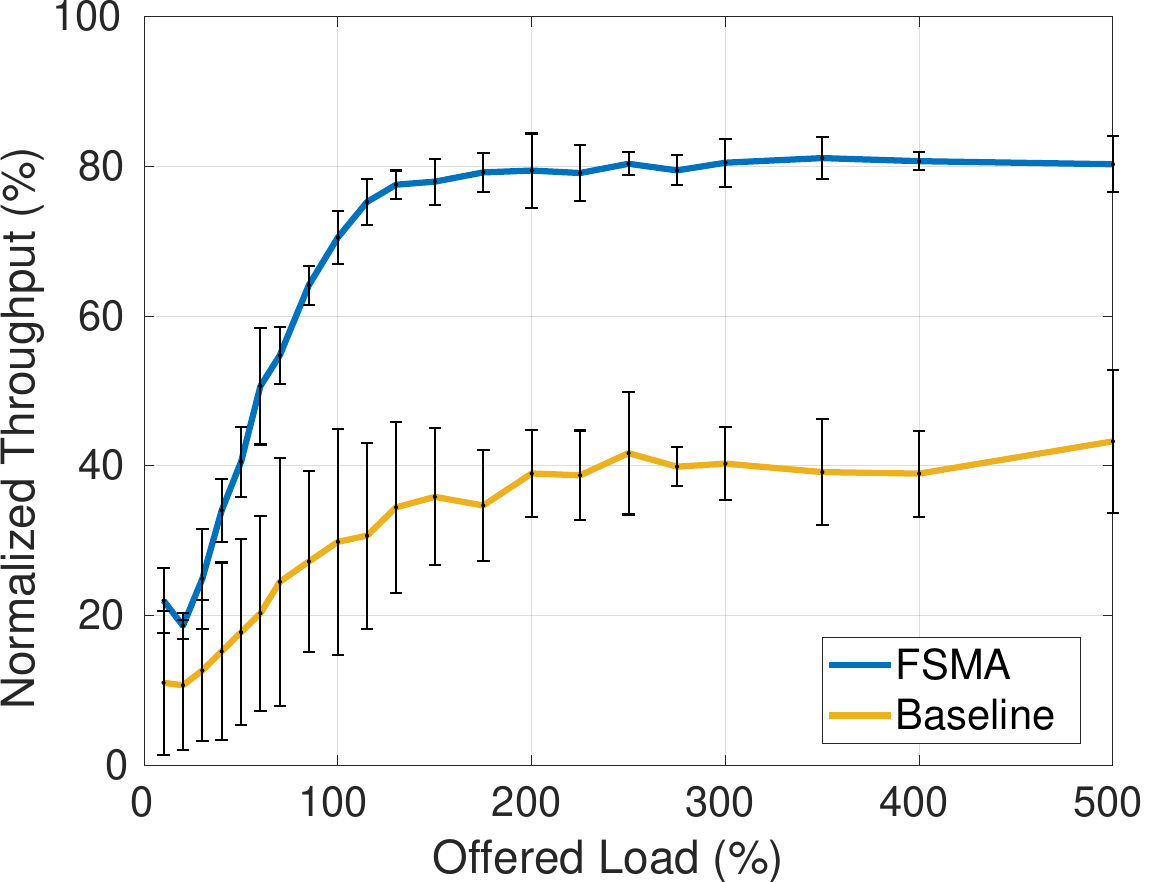}
         \caption{Throughput}
         \label{fig:eval_NormalizedThroughput}
     \end{subfigure}
     \hfill
     \begin{subfigure}[t]{0.23\textwidth}
         \centering
         \includegraphics[height=0.15\textheight]{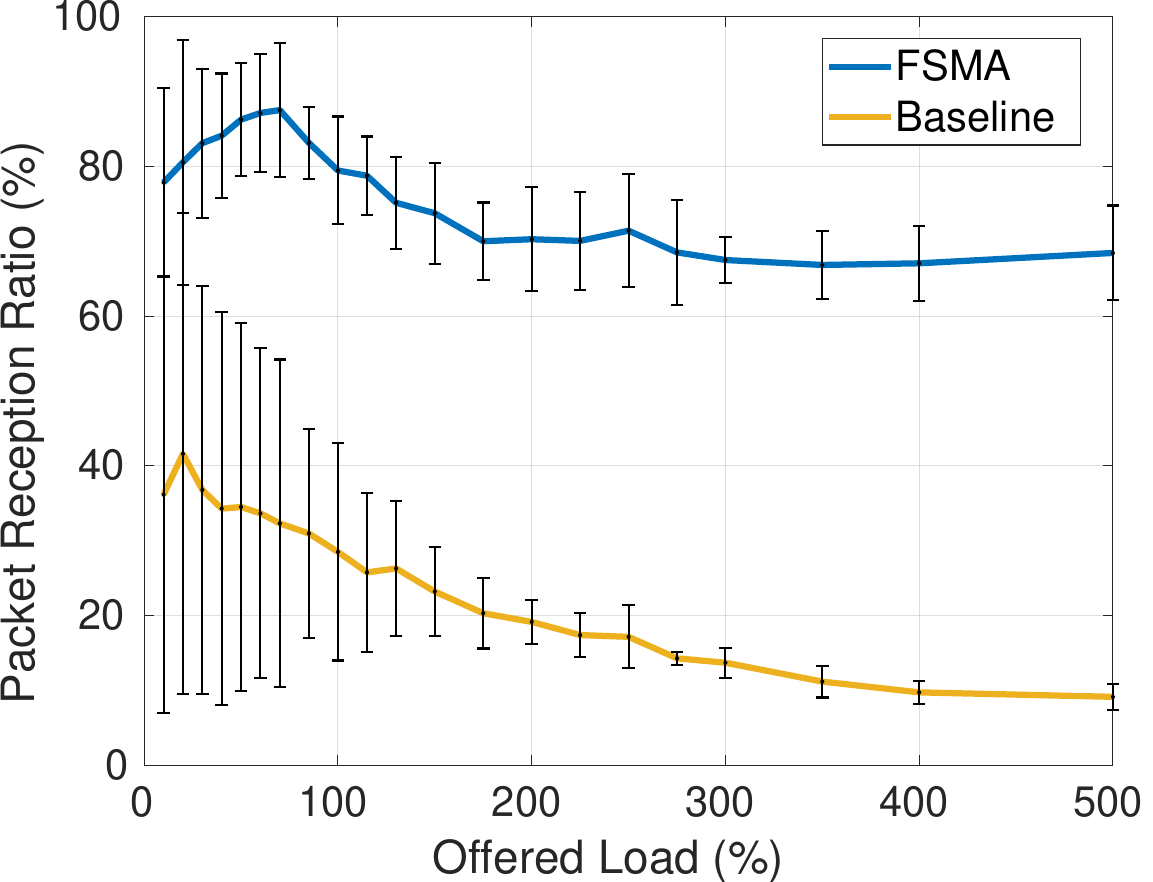}
         \caption{Packet reception ratio}
         \label{fig:eval_PacketReceptionRatio}
     \end{subfigure}
     \caption{Hardware experiments: static gateway.}
     \vspace{-0.01\textheight}
     \label{fig:eval_hardware_static}
\end{figure}

\noindent\textbf{Throughput:}
Uncoordinated access in baseline schemes caps throughput at 30\%
under mobility and 40\% in static conditions: collisions consume
channel time that carries no useful payload, and gateway mobility
adds link failures on transmissions that the baseline cannot suppress.
By reducing the collision window by 6$\times$--200$\times$ (Section~\ref{sec:collision_window}) and filtering out weak-link transmissions through SF-differential channel reciprocity (Section~\ref{sec:link_aware}), FSMA ensures that channel time carries decodable packets rather than collision noise, achieving 2$\times$ higher throughput in static settings and 2.5$\times$ under mobility
(Figures~\ref{fig:eval_throughput},~\ref{fig:eval_NormalizedThroughput}).

\begin{figure}[!t]
     \centering
     \begin{subfigure}[t]{0.23\textwidth}
         \centering
         \includegraphics[width=\textwidth]{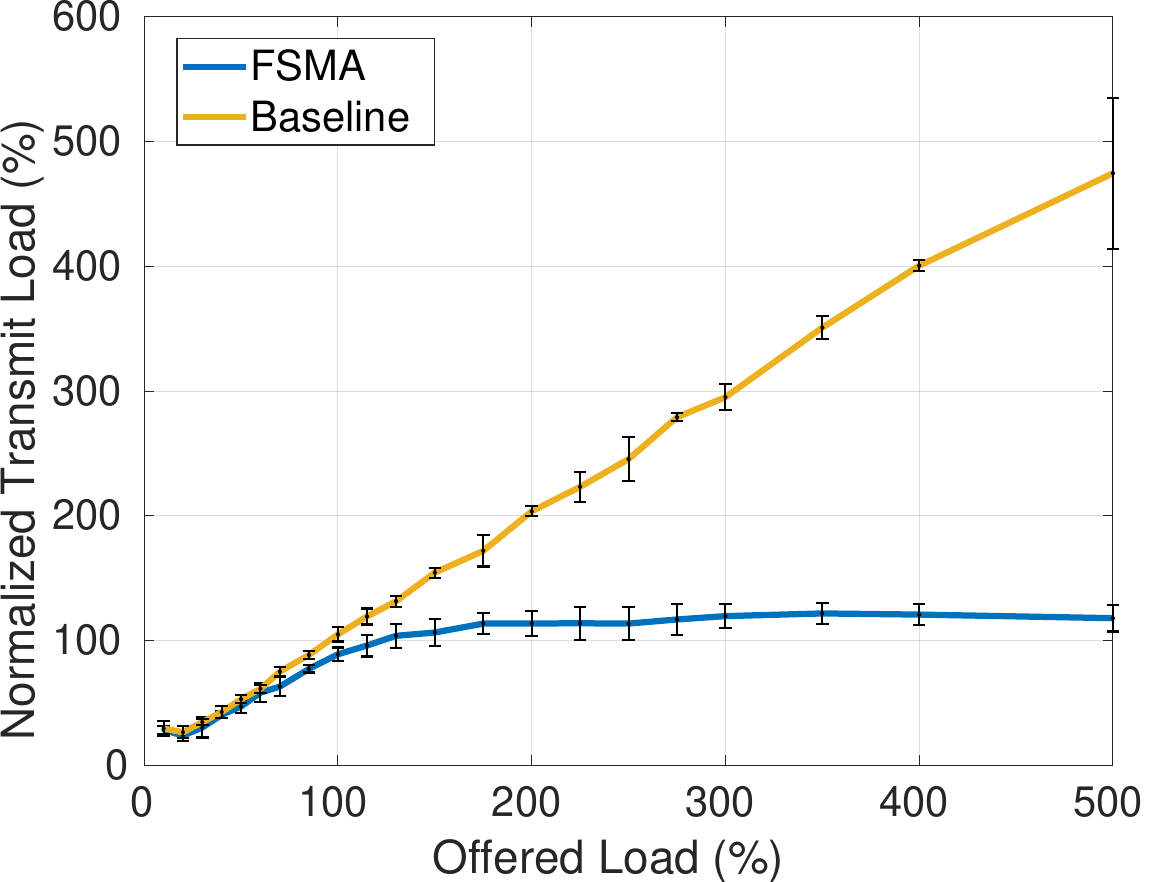}
         \caption{Normalized transmit load}
         \label{fig:eval_NormalizedTransmitLoad}
     \end{subfigure}
     \hfill
     \begin{subfigure}[t]{0.23\textwidth}
         \centering
         \includegraphics[width=\textwidth]{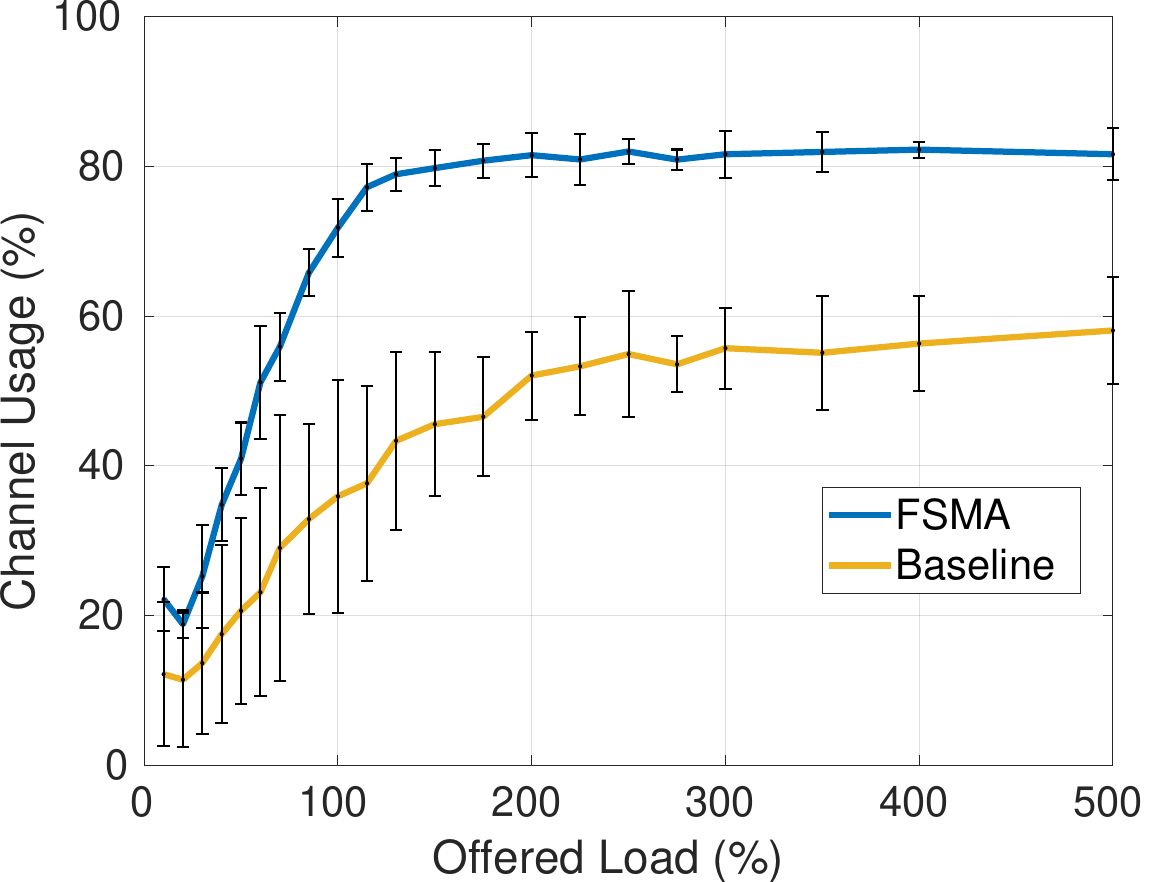}
         \caption{Channel usage}
         \label{fig:eval_ChannelUsage}
     \end{subfigure}
     \caption{FSMA features in a static setup. (a)~FSMA keeps
     transmit load near 100\%, preventing overload-induced collapse.
     (b)~FSMA maintains consistent 80\% channel occupancy.}
     \label{fig:eval_static_setup}
     \vspace{-0.03\textwidth}
\end{figure}

\noindent\textbf{Packet Reception Ratio:}
As offered load grows, baseline nodes transmit concurrently without
coordination, escalating collision rates. Once offered load exceeds
100\%, the transmit load grows linearly
(Figure~\ref{fig:eval_NormalizedTransmitLoad}) while the gateway's
decoding capacity is fixed, causing PRR to collapse. FSMA gates transmissions to a single FreeChirp window at a time, keeping the transmit load near network capacity regardless of offered load. This directly follows from the reduced collision window: because simultaneous transmissions are bounded to the sensing period, the gateway can always find a free window for the next FreeChirp. This
yields over 2.5$\times$ higher PRR at 100\% offered load and nearly
7$\times$ at 500\% load
(Figures~\ref{fig:eval_PRR},~\ref{fig:eval_PacketReceptionRatio}).

\noindent\textbf{Channel Usage:}
Baseline protocols waste channel capacity in two ways: collisions
render active transmissions undecodable, and idle gaps between
Poisson-distributed arrivals leave the channel unused. These effects
persist even above 100\% offered load, where the collision rate
outpaces any increase in transmit activity. FSMA coordinates access
via periodic \textit{FreeChirp} signals and randomized backoff,
filling the channel with decodable transmissions and sustaining
consistent 80\% occupancy across all load levels
(Figure~\ref{fig:eval_ChannelUsage}).

\vspace{-0.01\textheight}
\subsection{Large-Scale Simulations}
\label{sec:sim_eval}

\textit{NTNLoRa} simulations use real TLE data from a LEO satellite with
$\approx$10~minutes of visibility. Nodes are randomly distributed
across the coverage area (Figure~\ref{fig:sim_gateway_nodes_visual})
and the gateway follows the satellite pass, reproducing realistic
distance and link dynamics
(Figure~\ref{fig:sim_gateway_nodes_distances_plot}).


\noindent\textbf{FreeChirp Detection and Decoding Efficiency:}
The choice of FreeChirp SF governs the tradeoff between detection
reliability, coverage distance, and gateway energy. We evaluate
FreeChirp false negative rate and packet decoding efficiency at
distances up to 2000~km. As shown in
Figure~\ref{fig:eval_freechirp_detection}, no false positives occur
at any tested SF or distance. SF9 achieves zero missed detections up
to 2000~km with the lowest airtime among tested SFs, making it the
optimal FreeChirp configuration. For node uplinks, SF10 provides the
best balance of range and capacity. These results validate the SF9
FreeChirp and SF10 uplink configuration used throughout.

\begin{figure}[t]
 \centering
 \begin{subfigure}[t]{0.23\textwidth}
     \centering
     \includegraphics[width=\textwidth]{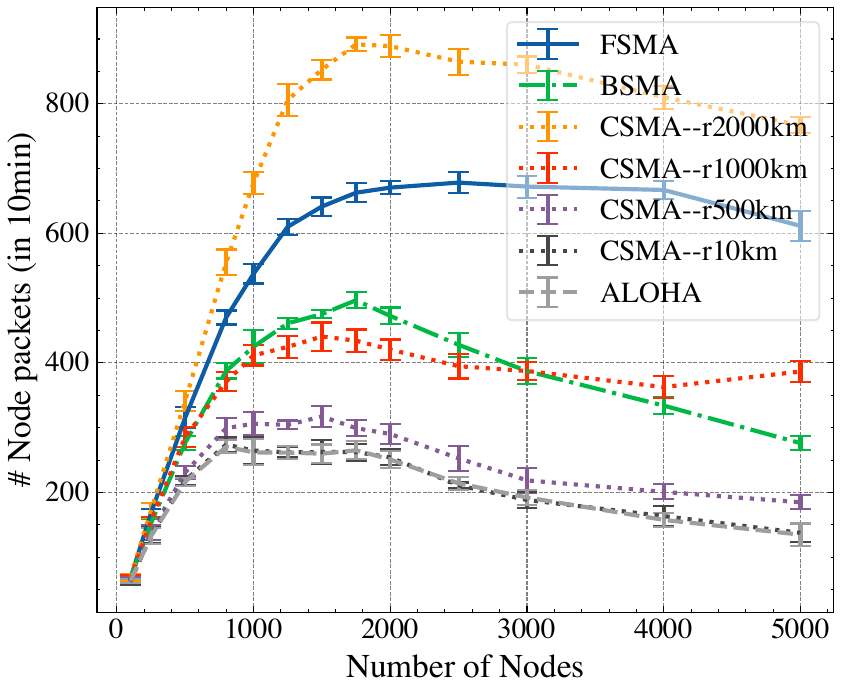}
     \vspace{-0.02\textheight}
     \caption{Received packets with nodes.}
     \label{fig:figures/sim_moving_received_packets_with_nodes.pdf}
 \end{subfigure}
 \hfill
 \begin{subfigure}[t]{0.23\textwidth}
     \centering
     \includegraphics[width=\textwidth]{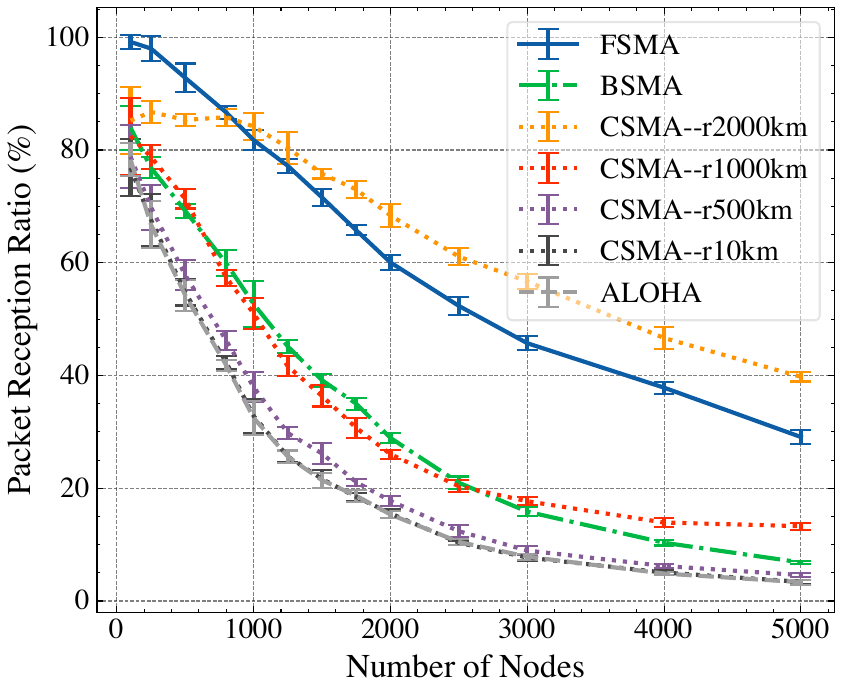}
     \vspace{-0.02\textheight}
     \caption{Packet reception ratio with nodes.}
     \label{fig:sim_moving_packet_reception_ratio_with_nodes}
 \end{subfigure}
    \caption{Moving gateway: FSMA vs. ALOHA, BSMA, and CSMA variants (10~km, 500~km, 1000~km, 2000~km oracle).
    \textcolor{blue}{Blue}: FSMA; \textcolor{green}{green}: BSMA;
    \textcolor{gray}{gray}: ALOHA.}
 \vspace{-0.01\textheight}
 \label{fig:simulator_moving_results}
\end{figure}

\noindent\textbf{Throughput and PRR at Scale:}
The moving scenario captures both collision avoidance and link-aware
access simultaneously. As shown in
Figure~\ref{fig:simulator_moving_results}, BSMA degrades sharply
because it has no mechanism to suppress transmissions from nodes with
poor links: weak packets occupy the channel but are not decoded. ALOHA
and CSMA are marginally better than expected because gateway mobility
reduces the instantaneous contending population, partially offsetting
collision effects. FSMA consistently outperforms all baselines,
achieving 3$\times$ higher received packets and 3--4$\times$
improvement in PRR. The early spike where FSMA slightly exceeds the
oracle reflects its link-awareness advantage: in low-contention
regimes where link variability dominates, FSMA prevents weak
transmissions that the oracle permits, improving decodable channel
usage.

\begin{figure}[t]
 \centering
 \begin{subfigure}[t]{0.23\textwidth}
     \centering
     \includegraphics[height=0.15\textheight]{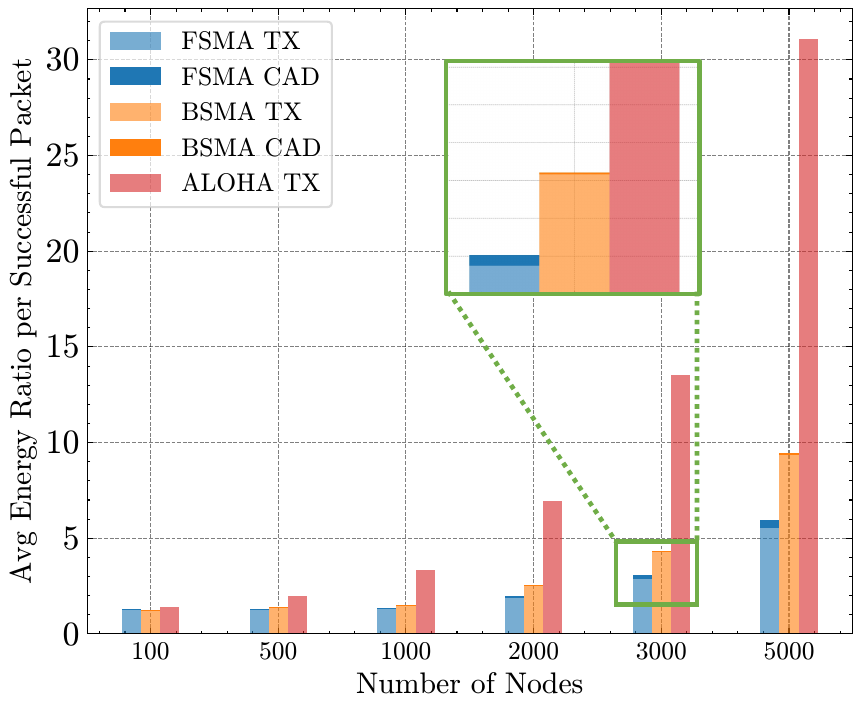}
     \vspace{-0.01\textheight}
     \caption{Node energy per successful packet.}
     \label{fig:sim_node_energy_ratio_per_packet}
 \end{subfigure}
 \hfill
 \begin{subfigure}[t]{0.235\textwidth}
     \centering
     \includegraphics[height=0.15\textheight]{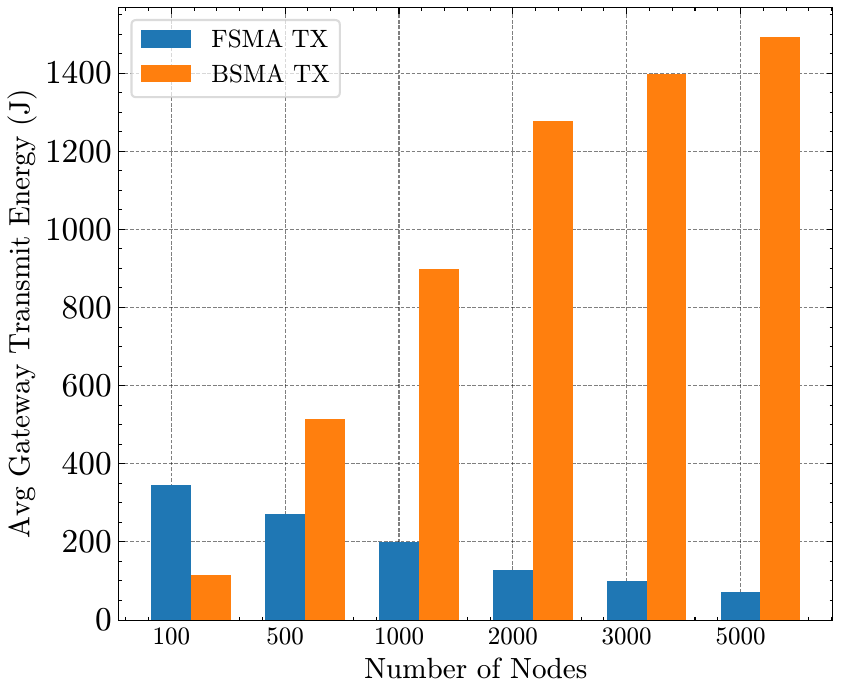}
      \vspace{-0.01\textheight}
     \caption{Gateway transmit energy overhead.}
     \label{fig:sim_gateway_energy_with_nodes}
 \end{subfigure}
 \caption{Energy efficiency under a moving gateway. BSMA only
 in~(b): ALOHA and CSMA impose no gateway-side overhead.}
 \vspace{-0.02\textheight}
 \label{fig:simulator_energy_results}
\end{figure}

\noindent\textbf{Energy Efficiency:}
At nodes, all protocols are comparable at low density. As scale
grows, ALOHA's collision rate forces repeated retransmissions,
sharply increasing energy per successful packet. BSMA and FSMA both
avoid retransmissions through collision mitigation, achieving similar
node-level efficiency. FSMA delivers up to 5$\times$ improvement over
ALOHA, including sensing overhead
(Figure~\ref{fig:sim_node_energy_ratio_per_packet}).
At the gateway, BSMA continuously transmits a busy tone when the
channel is occupied, incurring energy proportional to traffic load.
FSMA transmits a short \textit{FreeChirp} only when the channel is
free, reducing gateway energy consumption by up to 15$\times$ compared to BSMA
(Figure~\ref{fig:sim_gateway_energy_with_nodes}).

\begin{figure}[t!]
     \centering
     \begin{subfigure}[t]{0.23\textwidth}
         \centering
         \includegraphics[width=\textwidth]{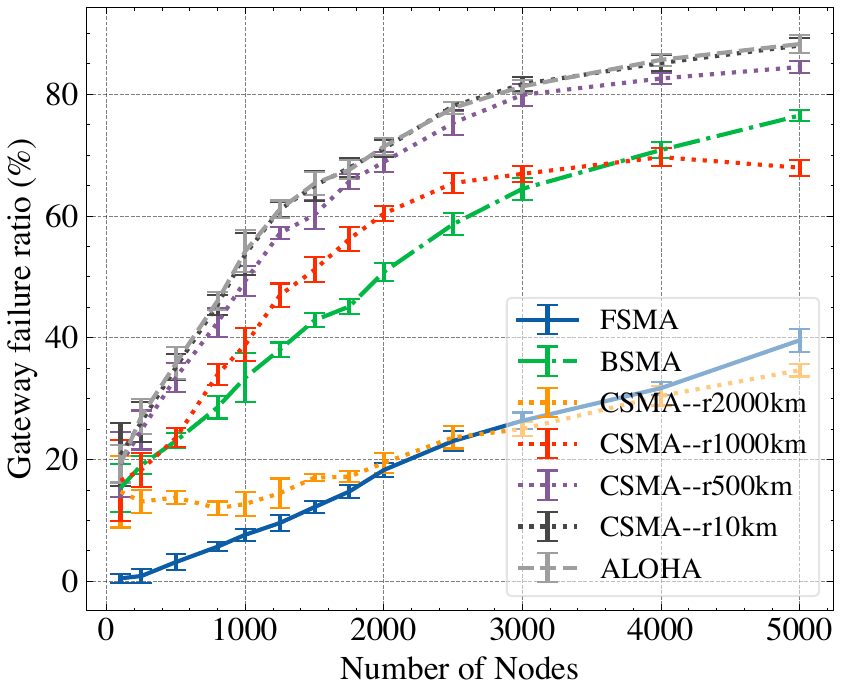}
         \vspace{-0.02\textheight}
         \caption{Gateway failure ratio (\%)}
         \label{fig:eval_gateway_failure_ratio_with_nodes}
     \end{subfigure}
     \hfill
     \begin{subfigure}[t]{0.23\textwidth}
         \centering
         \includegraphics[width=\textwidth]{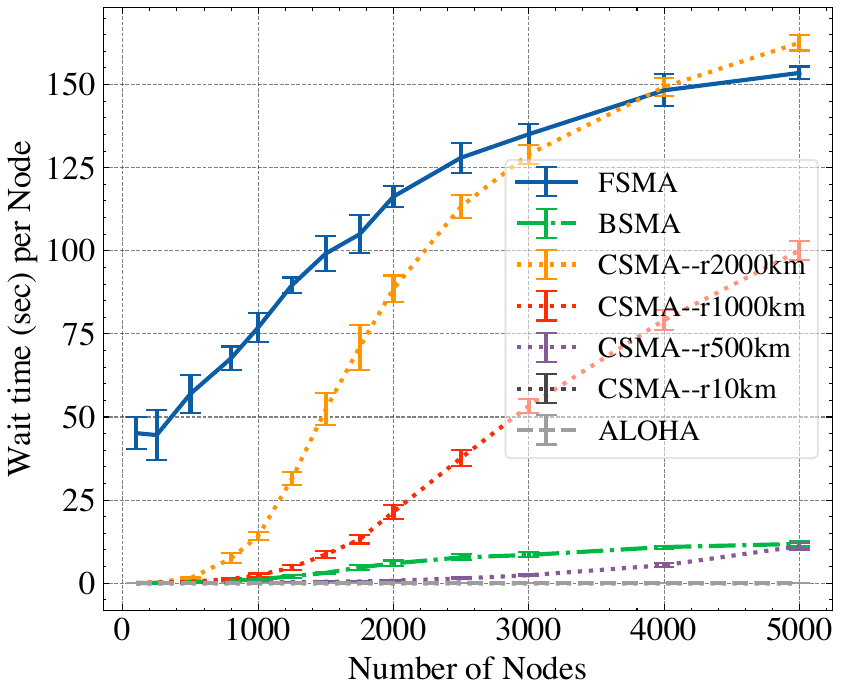}
        \vspace{-0.02\textheight}
         \caption{Node wait times}
         \label{fig:eval_wait_times_with_nodes}
     \end{subfigure}
     \caption{\algoname under a moving gateway. (a)~Gateway failure
     ratio: fraction of packets detected at the gateway but not
     decoded. (b)~Node wait times including CAD and backoff, averaging
     1--2.5~minutes within a satellite pass, well within
     acceptable bounds for NTN IoT.}
     \vspace{-0.02\textwidth}
     \label{fig:eval_failure_waittimes}
\end{figure}

\noindent\textbf{Gateway Failure Ratio:}
As node density increases, baseline protocols experience a rising
failure rate: overlapping arrivals exceed the SX127x 4-symbol locking
window (Section~\ref{sec:capture_effect}), causing the receiver to
lock onto incomplete signals. FSMA bounds arrival delay spread to
propagation delay differences, keeping colliding packets within the
capture window and enabling consistent decoding of the strongest
packet. FSMA reduces the gateway failure rate by up to 2.5$\times$
compared to baselines
(Figure~\ref{fig:eval_gateway_failure_ratio_with_nodes}).

\noindent\textbf{Node Wait Times:}
The per-node overhead from FSMA is packet wait time: the interval
between packet arrival and transmission, including CAD and backoff.
In static scenarios, wait times are short because backoff occurs only
for collision avoidance. Under gateway mobility, additional waiting
arises from link unavailability and low SNR, but
Figure~\ref{fig:eval_wait_times_with_nodes} shows that average wait
times stay within 1--2.5 minutes ($\approx$150~seconds) across all
node counts, well below the 10-minute satellite visibility window.
These delays are acceptable for NTN IoT applications including
precision agriculture, wildlife monitoring, and smart grids. Per
successful packet, FSMA's wait time substantially undercuts all
baselines by eliminating retransmissions from collisions and link
failures.

%% file: 6_conclusion.tex
\section{Limitations and Future Work}
\label{sec:limitations}

FSMA coordinates access within a single gateway's footprint; in
multi-gateway deployments, a node responding to one gateway's
FreeChirp may interfere with an adjacent gateway's reception,
addressable through orthogonal channel assignment across overlapping
gateways. Since FreeChirp is a bare upchirp, a rogue or cross-gateway
Chirp can trigger unnecessary transmissions, wasting node energy
without compromising payload security; custom beaconing with gateway
Identifiers mitigate this if needed. FSMA provides no explicit per-node 
fairness guarantee. Short-term fairness is not guaranteed, though fairness 
improves naturally as gateway mobility shifts coverage over time.

Three directions remain open for extending \algoname: multi-gateway coordination under
overlapping NTN coverage, concurrent decoding to recover multiple
simultaneous packets, and adaptive backoff strategies tailored to
NTN load dynamics and node location.

\section{Conclusion}
\label{sec:conclusion}

This paper presents FSMA, a synchronization-free, gateway-controlled
MAC protocol for LoRa-based NTN IoT networks. A single LoRa upchirp
(\shortname) simultaneously signals channel availability and filters
nodes with unreliable links through SF-differential channel
reciprocity, reducing the collision window by 6$\times$--200$\times$
without synchronization, handshakes, or hardware modifications.
Hardware experiments with 25 COTS nodes and a drone-mounted gateway
demonstrate 2$\times$ throughput gain, 2.5--5$\times$ better
packet reception ratio, and 5$\times$ improved energy efficiency.
\textit{NTNLoRa} simulations confirm scalability to 5000+ devices per satellite
pass, establishing FSMA as a practical and deployable solution for
next-gen NTN IoT.